\documentclass[twocolumn]{aastex6}

\usepackage{amssymb,amsmath,amsthm}
\usepackage{graphicx}
\usepackage{mathrsfs}
\usepackage{booktabs}
\usepackage{multirow}
\usepackage{empheq}
\usepackage{verbatim}

\usepackage{natbib}
\shortauthors{Pesce et al.}

\makeatletter
\def\env@matrix{\hskip -\arraycolsep
  \let\@ifnextchar\new@ifnextchar
  \array{*{\c@MaxMatrixCols}c}}
\makeatother

\begin{document}

\title{Measuring supermassive black hole peculiar motion using H$_2$O megamasers}

\author{D. W. Pesce}
\affil{Department of Astronomy, University of Virginia, 530 McCormick Road, Charlottesville, VA 22904, USA}
\affil{National Radio Astronomy Observatory, 520 Edgemont Road, Charlottesville, VA 22903, USA}
\email{dpesce@virginia.edu}
\and
\author{J. A. Braatz}
\affil{National Radio Astronomy Observatory, 520 Edgemont Road, Charlottesville, VA 22903, USA}
\and
\author{J. J. Condon}
\affil{National Radio Astronomy Observatory, 520 Edgemont Road, Charlottesville, VA 22903, USA}
\and
\author{J. E. Greene}
\affil{Department of Astrophysics, Princeton University, Princeton, NJ, USA}

\begin{abstract}
H$_2$O megamasers residing in the accretion disks of active galactic nuclei (AGN) exhibit Keplerian rotation about the central supermassive black hole (SMBH).  Such disk maser systems are excellent tools for diagnosing the kinematic status of the SMBH, and they currently provide the only direct and unambiguous measure of SMBH velocities outside of the Milky Way.  We have measured the galaxy recession velocities for a sample of 10 maser disk systems using a combination of spatially resolved HI disk modeling, spatially integrated HI profile fitting, and optical spectral line and continuum fitting.  In comparing the SMBH velocities to those of their host galaxies, we find two (out of 10) systems -- J0437+2456 and NGC 6264 -- for which the SMBH and galaxy velocities show a statistically significant ($>$3$\sigma$) difference.  For NGC 6264 the apparent velocity offset can likely be explained by ionized gas motion within the host galaxy (e.g., from AGN-driven shocks).  The velocity measurements for J0437+2456, however, imply a SMBH peculiar velocity of $69.6 \pm 12.7$~km~s$^{-1}$ (5.5$\sigma$).  We thus consider J0437+2456 to be a promising candidate for hosting either a recoiling or binary SMBH, though additional observations are necessary to exclude the possibility of a systematic offset between the galactic recession velocity and that measured using the optical spectrum.
\end{abstract}


\section{Introduction}

A supermassive black hole (SMBH) in kinetic equilibrium with its surrounding stellar environment will be nearly motionless ($v \ll 1$ km~s$^{-1}$; \citealt{2007AJ....133..553M}) with respect to the system barycenter.  Any larger relative motions can result from several mechanisms:

\begin{enumerate}
	\item \underline{SMBH binary orbital motion}: Even a relatively low-mass ($\sim$$10^7$ M$_{\odot}$), wide-separation (hundreds of parsecs) SMBH binary can exhibit orbital motions exceeding 10 km~s$^{-1}$.  By the time dynamical friction has ceased being efficient and the binary has possibly stalled at the ``final parsec" (\citealt{1980Natur.287..307B}; \citealt{2003ApJ...596..860M}), the orbital velocities will be well in excess of 100 km~s$^{-1}$.
	\item \underline{Gravitational wave recoil}: The merging of two SMBHs results in the anisotropic radiation (in the form of gravitational waves) of not only mass and angular momentum from the system, but also linear momentum \citep{1973ApJ...183..657B}.  Depending on various details of the precursor systems (e.g., spin configuration and mass ratio), the resulting recoil of the remnant SMBH can easily be several hundred km~s$^{-1}$, and in some cases might even reach thousands of km~s$^{-1}$ (\citealt{2004ApJ...607L...5F}; \citealt{2007PhRvL..98w1102C}).  The largest of these kicks would eject the SMBH from most stellar systems, but SMBHs with kicks not exceeding the escape velocity will experience orbital decay from dynamical friction with the surrounding stars, gas, and dark matter, causing them to undergo a damped oscillation about the center of the galaxy \citep{2004ApJ...607L...9M}.
	\item \underline{Ongoing galaxy merger}: As two galaxies merge, the SMBH from one galaxy will not initially be in equilibrium with the stellar system from the other galaxy (e.g., \citealt{2014ApJ...789..112C}).  In this case, we might observe relative motion between the two resulting from our inability to observationally disentangle the velocity contributions from the two dynamically distinct stellar systems.
	\item \underline{Massive perturbers}: The presence of massive objects (e.g., star clusters, molecular clouds, etc.) in the SMBH's environment will cause the equilibrium velocity dispersion (``gravitational Brownian motion") to increase over that expected from just the stellar population alone \citep{2007AJ....133..553M}.  However, unless the massive objects constitute a non-negligible fraction of the environment by mass, the increase in SMBH velocity dispersion (which scales approximately as the square root of the characteristic perturber mass) will be unnoticeable.  A more pronounced perturbation might occur when a massive perturber (e.g., molecular cloud) passes very close to the SMBH, an interaction which simulations have tentatively shown may result in large kicks (e.g., \citealt{2013MNRAS.434..606G}).
	\item \underline{Jet-powered rocket}: If an AGN's jets are intrinsically asymmetric, the resulting net acceleration can propel the SMBH to observable displacements and velocities (\citealt{1982IAUS...97..475S}).  This mechanism, if prevalent, could be weakened by ``flip-flop" instabilities that cause the more strongly emitting jet component to repeatedly alternate sides \citep{1984ApJ...279...74R}.  However, we note that the appearance of one-sided jets can often be explained by relativistic beaming without the need for any intrinsic asymmetry (e.g., \citealt{1983Natur.303..779E}).
	\item \underline{Three-body scattering}: If a galaxy merger occurs where one of the galaxies contains a binary SMBH system, the SMBH from the other galaxy can experience strong three-body scattering off of the binary \citep{2006ApJ...638L..75H}.  Such a scenario potentially explains the ``naked" quasar HE 0450-2958 \citep{2005Natur.437..381M}, which is observed to be displaced by $\sim$7 kpc from a galaxy that appears to have undergone a recent merger (though other scenarios are possible; see, e.g., \citealt{2007ApJ...658..107K}).
\end{enumerate}

For the majority of SMBH systems, we consider only the first three of these mechanisms to be likely causes of sizable (i.e., several km~s$^{-1}$ or larger) ``peculiar velocities": i.e., motion of the SMBH relative to the stellar system that significantly exceeds what would be expected in equilibrium.  Observational efforts to identify SMBH peculiar motion primarily attempt to detect one of three predicted signatures: (1) velocity offsets between the SMBH and the galactic barycenter (e.g., \citealt{2009ApJ...698..956C}, \citealt{2009ApJ...705L..76W}, \citealt{2016ApJ...824..122K}), (2) positional offsets between the SMBH and the galactic barycenter (e.g., \citealt{2003ApJ...582L..15K}, \citealt{2013ApJ...762..110L}, \citealt{2016ApJ...829...37B}), or (3) gravitational wave emission (e.g., \citealt{2014ApJ...794..141A}, \citealt{2014MNRAS.444.3709Z}, \citealt{2016MNRAS.455.1665B}).  In this paper we focus on the first of these signatures.

To spectroscopically identify SMBH peculiar motion, one can either compare the velocities of the SMBH and surrounding system at a single epoch or monitor the SMBH velocity over time.  Both methods require some observational measure of the SMBH velocity, and the first method also requires an observational measure of the galactic velocity.  A spectroscopic measurement of the SMBH velocity can only be made if the black hole has some emitting material that shares its motion (e.g., a gravitationally bound accretion disk), effectively limiting such measurements to active galactic nuclei (AGN).

Past efforts to make velocity measurements of SMBHs have typically used optical spectra, either decomposing the emission lines into broad and narrow components (e.g., \citealt{2016ApJ...824..122K}) or looking for shifts in the broad line centroids with time (e.g., \citealt{2013ApJ...777...44J}).  The idea here is that the broad line region (BLR) traces gas in the immediate vicinity of the SMBH while the narrow line region (NLR) is thought to share the host galaxy's recession velocity.  Any discrepancy between the BLR and NLR central velocities at the same epoch, or any change in the BLR velocity with time, could then be an indication of SMBH peculiar motion.  There are many examples in the literature of this class of search (e.g., \citealt{2007ApJ...666L..13B}, \citealt{2009Natur.458...53B}, \citealt{2012ApJS..201...23E}, \citealt{2017ApJ...834..129W}).

Attempts to measure SMBH peculiar motions using optical spectra suffer from several difficulties.  Single-epoch measurements of SMBH peculiar motion using optical lines are hindered by the broad and blended nature of the emission lines.  The statistical uncertainty in any relative velocity measurement will be an increasing function of the widths of the lines used, thereby limiting the precision of such measurements.  Differential reddening and flux asymmetries also heavily impact the accuracy of any velocity reconstruction made using broad lines (e.g., \citealt{2002AJ....124....1R}).  Systematic velocity offsets between the BLR and NLR (as traced using, e.g., [OIII]) are observed to be fairly common (e.g., \citealt{2005AJ....130..381B}, \citealt{2009MNRAS.394L..16M}, \citealt{2012ApJ...756...51L}), and are likely produced by NLR dynamics (e.g., AGN-driven outflows) rather than by SMBH peculiar motion.  \citealt{2005AJ....130..381B} in particular claims that the [O II] $\lambda$3727, [N II] $\lambda$6548, $\lambda$6584, and [S II] $\lambda$6716, $\lambda$6731 lines likely trace the systemic velocity of the galaxy, while the [O III] $\lambda$4959, $\lambda$5007 lines are often ($\sim$50\% of the time) systematically blueshifted by tens to hundreds of km~s$^{-1}$ with respect to lower-ionization lines.  Multiple-epoch measurements are further plagued by emission lines that depend sensitively on the spatial scales probed by the spectrum (see, e.g., \citealt{2006ApJ...636..654R}), and which can vary across epochs either intrinsically (e.g., \citealt{2017MNRAS.468.1683R}) or when different facilities or fiber/slit placements are used.

In this paper we present a technique for measuring SMBH peculiar motions with unprecedented precision.  H$_2$O megamasers residing in the accretion disks of AGN provide an excellent means of diagnosing
the kinematic status of a SMBH.  By tracing the Keplerian rotation curves of megamaser disks only tenths of a parsec from the nucleus, and well within the SMBH ``sphere of influence," one can measure not only the mass of the SMBH but also the absolute on-sky position and line-of-sight velocity of the dynamic center (e.g., \citealt{2011ApJ...727...20K}, \citealt{2017ApJ...834...52G}).  Such measurements require a very long baseline interferometric (VLBI) map to spatially resolve the maser system.  Coupling high-sensitivity VLBI maps with multi-year spectral monitoring further enables ``full disk" modeling, yielding measurements of the 3-dimensional disk geometry, black hole mass, and distance to the system.  The megamaser technique results in very precise measurements of several relevant quantities, with uncertainties generally $\lesssim$1 mas in the SMBH absolute position and $\lesssim$2 km~s$^{-1}$ in its velocity (e.g., \citealt{2013ApJ...767..154R}, \citealt{2013ApJ...767..155K}, \citealt{2013ApJ...775...13H}, \citealt{2015ApJ...800...26K}, \citealt{2016ApJ...817..128G}).

In principle, any method that uses orbital analysis to determine the mass of a SMBH (e.g., CO gas disks with ALMA; \citealt{2016ApJ...822L..28B}) will also necessarily constrain its velocity.  However, in extragalactic environments AGN disk masers are currently the only tools available that directly probe the SMBH gravitational sphere of influence without needing to account for various contaminating effects such as foreground reddening/absorption or the need to model the stellar distribution or gas turbulence profile.  That is, \textit{AGN disk masers exhibiting Keplerian rotation about the central SMBH currently provide the only direct and unambiguous measure of its velocity}, independent of the surrounding material.

This paper is organized as follows.  In \S\ref{ObservationsAndReduction} we present both new and archival neutral hydrogen (HI) observations from the Karl G. Jansky Very Large Array (VLA), and we describe the data reduction and imaging procedures.  In \S\ref{GalaxyRecessionVelocities} we outline the analyses performed on the HI data to extract galaxy recession velocity measurements, and in \S\ref{MeasuringSMBHVelocities} we describe the analyses for measuring the SMBH velocities using maser data.  \S\ref{Results} gives a detailed discussion of the results for each galaxy in our sample.  Unless otherwise specified, all velocities referenced in this work use the optical definition in the barycentric reference frame.

\section{Observations and data reduction} \label{ObservationsAndReduction}

The galaxies analyzed in this paper were selected because they have published VLBI observations of H$_2$O megamaser emission in Keplerian rotation around the SMBH.  We present new VLA HI observations of 7 of these galaxies (from NRAO project 16A-238), out of which we detected 4.  We have also retrieved VLA HI observations of NGC 1194, UGC 3789, Mrk 1419, and NGC 4258 from the NRAO archives; we note that the archival observations of NGC 1194, UGC 3789, and Mrk 1419 have been previously published in \cite{2013ApJ...778...47S}.  We do not present any new VLBI observations in this paper, but we have re-analyzed the VLBI data from \cite{2011ApJ...727...20K} to fit rotation curves to the maser disks in NGC 2273 and NGC 1194 (see \S\ref{FittingRotationCurves}); see that paper for details on the data reduction.  For the H$_2$O maser system in NGC 4258, we use the fitting results from \cite{2013ApJ...775...13H}.  Ultimately, only the HI observations of NGC 2273, NGC 1194, and NGC 4258 proved conducive to the tilted-ring model we sought to apply for measuring galaxy recession velocities (see \S\ref{GalaxyRecessionVelocities}).

Information about the VLA observations is listed in Table \ref{tab:Observations}.  Nearly all observations were taken with the VLA in C configuration, with the one exception being the archival D configuration observations of NGC 4258.  For observations from project 16A-238, the correlator was configured with a single 16~MHz spectral window centered on the HI 21~cm spin-flip transition (L-band).  We observed in dual circular polarization using 4096 channels across the bandwidth, corresponding to a channel size of $\sim$3.91~kHz ($\sim$0.85~km~s$^{-1}$).  For the observations of NGC 4258, two 3~MHz spectral windows covered the HI line, with 0.68~MHz of overlap at the center.  The observations were carried out in dual polarization, using 31 contiguous 97.7~kHz spectral channels ($\sim$20 km~s$^{-1}$) across the bandwidth.  Details on the observational setup for NGC 1194, UGC 3789, and Mrk 1419 are reported in \cite{2013ApJ...778...47S}.

We reduced all VLA data using standard procedures with the Common Astronomy Software Applications (CASA) package\footnote{https://casa.nrao.edu/}.  After correcting for antenna positions and atmospheric opacity, we solved for delay and phase solutions on the flux calibrator (which also doubled as our bandpass calibrator).  With these solutions applied to the flux calibrator we then used it to obtain the bandpass shape, applied the bandpass correction to all calibrators, and solved for the gains and fluxes.  All solutions were applied to the target, after which we typically performed a round of flagging and iterated on the calibration once more before ultimately splitting out the calibrated science target.  Radio frequency interference (RFI) was the most common reason for flagged data; we sliced the observations across time, frequency, polarization, and baseline to isolate and excise RFI.  Prior to imaging, we performed continuum subtraction on the UV data.  We used natural UV weighting when imaging the data cubes with CLEAN, and we subsequently corrected for primary beam attenuation before performing any of the visualization described in the next section.

\floattable
\begin{deluxetable}{lcccccccccc}
\tablecolumns{11}
\tabletypesize{\tiny}
\tablecaption{VLA observations\label{tab:Observations}}
\tablehead{	&	\colhead{R.A.}	&	\colhead{Decl.}	&		&	\colhead{Gain}	&	\colhead{Flux}	&	\colhead{$t_{\text{tot}}$}	&	\colhead{Synthesized beam}	&	\colhead{Channel size}	&	\colhead{Noise}	&	\colhead{Peak intensity} \\
\colhead{Galaxy}	&	\colhead{(J2000)}	&	\colhead{(J2000)}	&	\colhead{Date(s)}	&	\colhead{calibrator}	&	\colhead{calibrator}	&	\colhead{(min.)}	&	\colhead{($'' \times ''$, deg.)}	&	\colhead{(kHz)}	&	\colhead{(mJy~beam$^{-1}$)}	&	\colhead{(mJy~beam$^{-1}$)}}
\startdata
NGC 1194			&	03:03:49.1	&	$-$01:06:13		&	2010 Oct 08									&	J0323$+$0534	&	3C 48		&	212	&	$22.2 \times 16.8$, $-0.67$	&	15.625	&	2.1		&	4.2			\\
J0437+2456		&	04:37:03.7	&	$+$24:56:07		&	2016 Mar 17, Mar 20, Mar 24	&	J0431$+$2037	&	3C 138	&	276	&	$18.0 \times 16.9$, $-34.1$	&	3.906		&	2.7		&	\ldots	\\
NGC 2273			&	06:50:08.6	&	$+$60:50:45		&	2016 Feb 28, Mar 13					&	J0614$+$6046	&	3C 147	&	297	&	$27.1 \times 17.4$, $-81.8$	&	3.906		&	2.4		&	9.7			\\
ESO 558-G009	&	07:04:21.0	&	$-$21:35:19		&	2016 Apr 24									&	J0706$-$2311	&	3C 147	&	92	&	$35.8 \times 15.4$, $-24.7$	&	3.906		&	4.5		&	2.3			\\
UGC 3789			&	07:19:30.9	&	$+$59:21:18		&	2010 Oct 07									&	J0614$+$6046	&	3C 147	&	198	&	$23.6 \times 16.9$, $77.9$	&	15.625	&	1.6		&	2.6			\\
Mrk 1419			&	09:40:36.4	&	$+$03:34:37		&	2010 Nov 19									&	J0943$+$0819	&	3C 286	&	210	&	$19.4 \times 16.3$, $-30.1$	&	15.625	&	0.96	&	1.0			\\
NGC 4258			&	12:18:57.5	&	$+$47:18:14		&	1994 Jan 02, Jan 03					&	J1150$+$497		&	3C 48		&	37	&	$74.5 \times 54.3$, $-88.2$	&	97.656	&	0.80	&	198			\\
CGCG 074-064	&	14:03:04.4	&	$+$08:56:51		&	2016 Mar 29, Apr 08, Apr 09	&	J1347$+$1217	&	3C 286	&	280	&	$23.7 \times 17.7$, $44.2$	&	3.906		&	2.1		&	\ldots	\\
NGC 5765b			&	14:50:51.5	&	$+$05:06:52		&	2016 Mar 07, Mar 08					&	J1445$+$0958	&	3C 286	&	297	&	$20.6 \times 17.3$, $10.4$	&	3.906		&	2.1		&	1.4			\\
NGC 6264			&	16:57:16.1	&	$+$27:50:59		&	2016 Mar 15, Mar 17					&	J1613$+$3412	&	3C 295	&	264	&	$22.3 \times 17.4$, $-67.7$	&	3.906		&	2.7		&	\ldots	\\
NGC 6323			&	17:13:18.1	&	$+$43:46:57		&	2016 Mar 11, Mar 13					&	J1635$+$3808	&	3C 295	&	278	&	$19.2 \times 17.6$, $82.0$	&	3.906		&	2.2		&	0.73		\\
\enddata
\tablecomments{Information about the VLA observations.  The listed coordinates give the phase center supplied to the correlator.  The total on-source observing time is denoted $t_{\text{tot}}$), and the beam position angle is given in degrees east of north.  The rms noise level achieved is quoted per spectral channel.  For sources where HI was detected, the peak HI line intensity (as determined by the Gaussian fitting procedure described in \S\ref{DataVisualization}) is given in the last column.}
\end{deluxetable}

\subsection{Data visualization and masking} \label{DataVisualization}

Rather than displaying the resulting image cubes using moment maps, which tend to be dominated by noise features in low signal-to-noise data, we opted for a parametric fitting approach similar to that employed by \cite{2013ApJ...778...47S}.  For every spatial pixel in each image cube, we used a Markov Chain Monte Carlo (MCMC) code to fit a Gaussian line profile to the spectrum extracted at that location.  The resulting posterior distribution allows us to associate a best-fit line amplitude, velocity centroid, and velocity dispersion with every spatial pixel in the image cube, along with the corresponding uncertainty (determined by the width of the marginalized posterior) in each parameter.  These parameter values then enable us to construct the equivalent of moment maps using the fitted Gaussian profiles rather than the (noisy) data, resulting in considerable aesthetic improvement.  More practically, this parametric fitting technique is also more sensitive to low-amplitude line emission than the ``by-eye" detection methods that moment maps are often used for.  Our image cubes for ESO 558-G009 and NGC 6323, for instance, do not show any obvious line emission in moment maps created using CASA, but the Gaussian fits are able to extract the extant signal without spectral or spatial averaging.

The parametric fitting technique also provides a natural way to mask the data when creating different moment maps (and equivalents).  Rather than using a signal-to-noise cut, which for interferometric data inevitably results in either a ``splotchy" image or (if a high enough signal-to-noise cut is used) ends up masking out some of the real signal, we instead used the uncertainties in the fitted line profiles to determine which pixels are trustworthy.  For each image cube we performed a 4$\times$4-pixel spatial smoothing of each fitted parameter, and we retained only those spatial pixels for which all smoothed parameter values exceeded 3 times the corresponding 1$\sigma$ uncertainty.  These masks were then applied to the original (unsmoothed) image maps.

\begin{figure*}[t]
	\centering
		\includegraphics[width=1.00\textwidth]{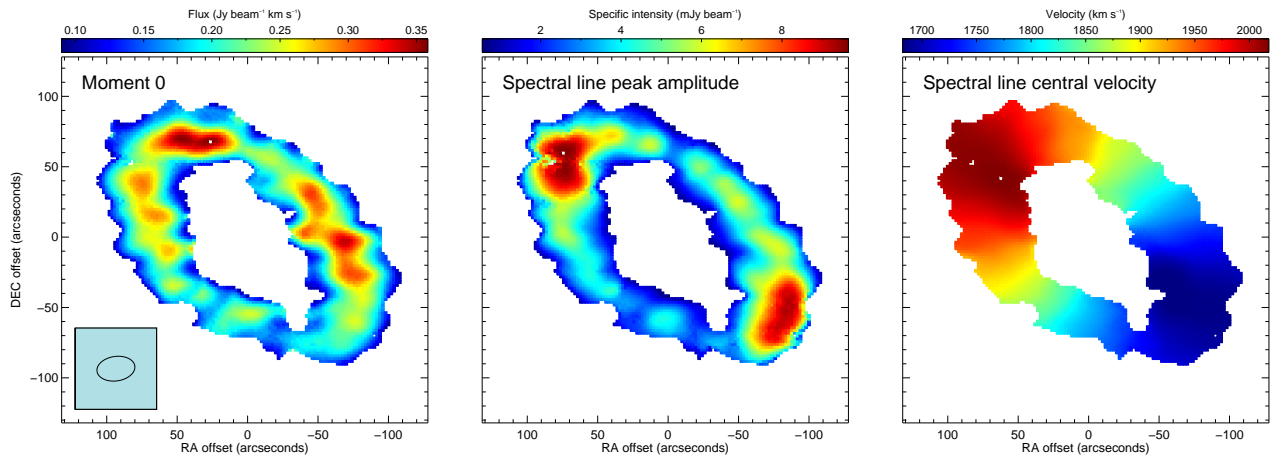}
	\caption{Moment 0 map (left), spectral line peak amplitude map (center), and velocity map (right) of the HI in NGC 2273, created and masked using the procedure described in \S\ref{DataVisualization}.  The coordinate axes mark the offset in right ascension and declination from the phase center of the observations (see Table \ref{tab:Observations}), and the half-power beam shape is shown in the bottom left-hand corner of the leftmost plot.}
	\label{fig:NGC2273_HI_maps}
\end{figure*}

\begin{figure*}[t]
	\centering
		\includegraphics[width=1.00\textwidth]{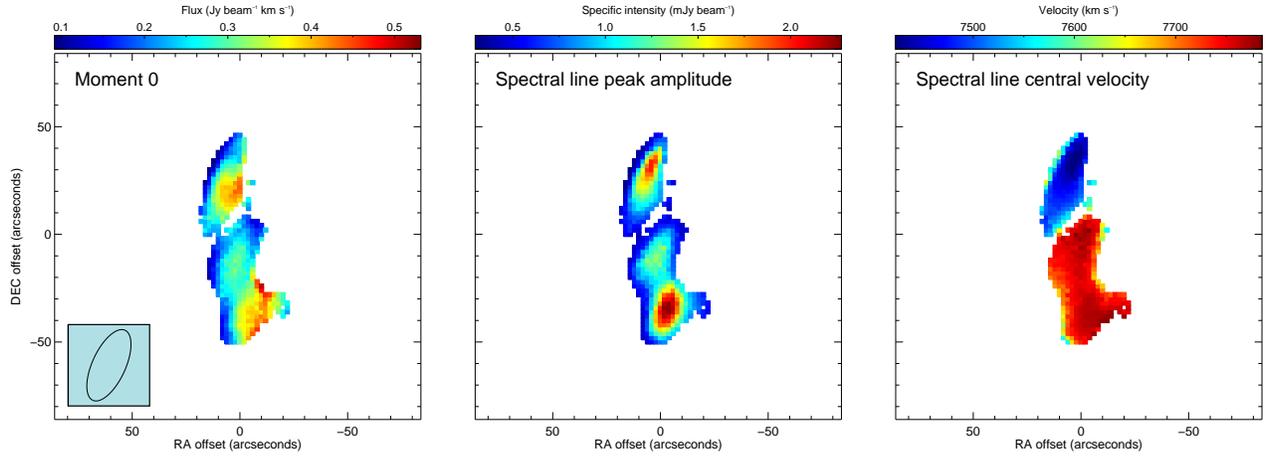}
	\caption{Same as Figure \ref{fig:NGC2273_HI_maps}, but for ESO 558-G009.}
	\label{fig:ESO558-G009_HI_maps}
\end{figure*}

\begin{figure*}[t]
	\centering
		\includegraphics[width=1.00\textwidth]{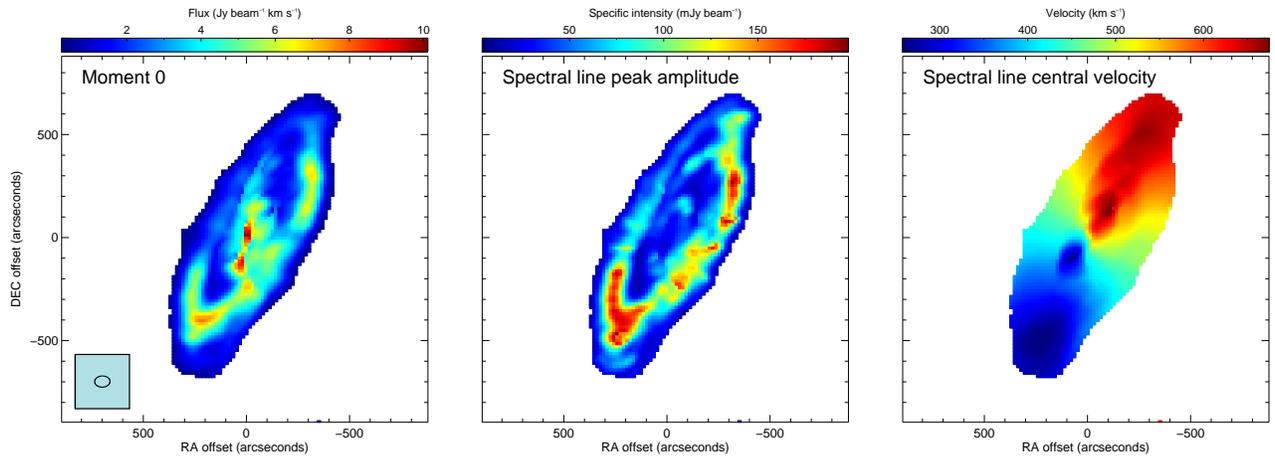}
	\caption{Same as Figure \ref{fig:NGC2273_HI_maps}, but for NGC 4258.}
	\label{fig:NGC4258_HI_maps}
\end{figure*}

\begin{figure*}[t]
	\centering
		\includegraphics[width=1.00\textwidth]{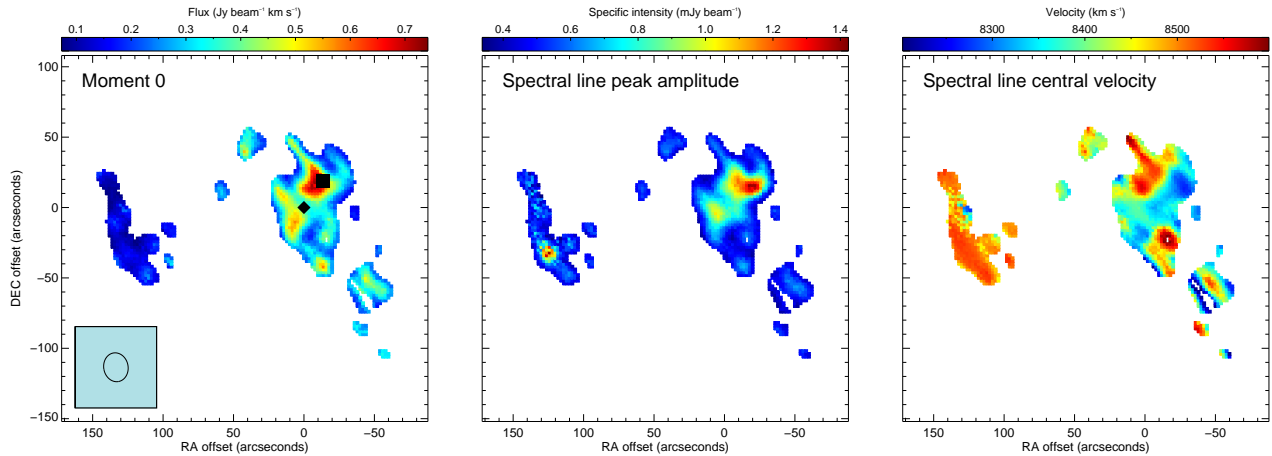}
	\caption{Same as Figure \ref{fig:NGC2273_HI_maps}, but for NGC 5765b.  The optical center of NGC 5765b is marked in the leftmost panel with a black diamond, while the optical center of NGC 5765a is marked with a black square.  The positions of both galaxies have been taken from NED.}
	\label{fig:NGC5765b_HI_maps}
\end{figure*}

\begin{figure*}[t]
	\centering
		\includegraphics[width=1.00\textwidth]{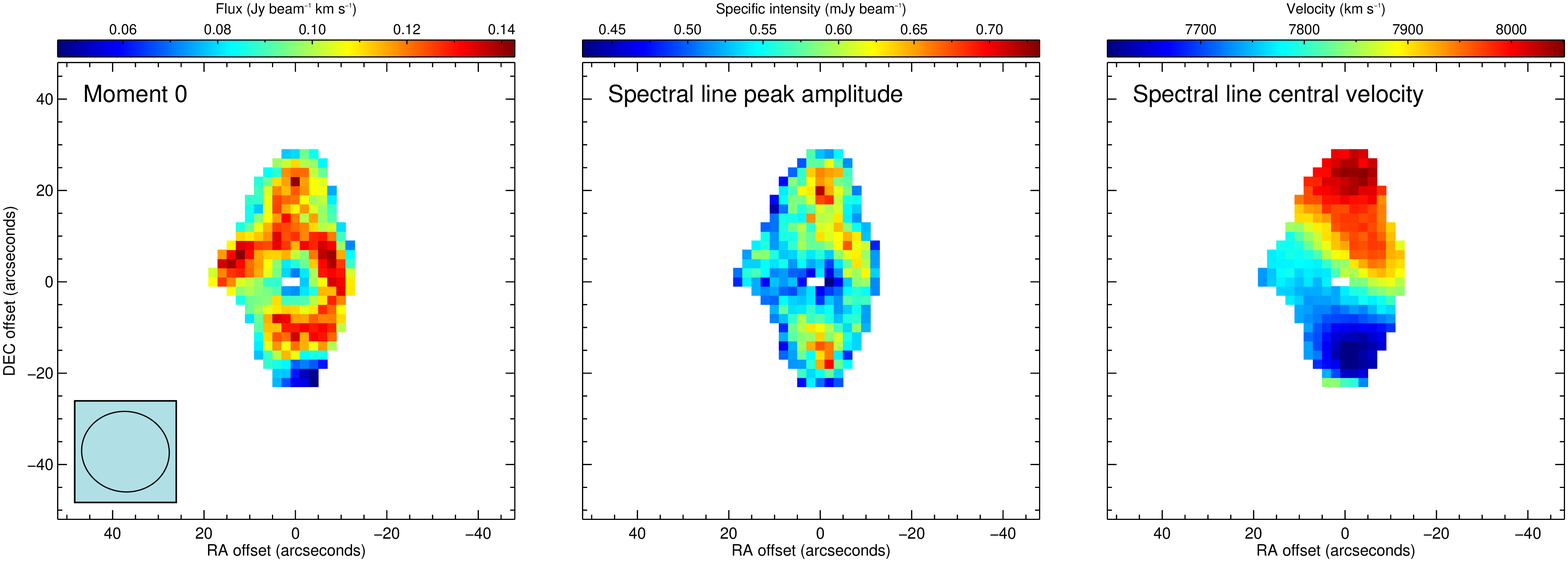}
	\caption{Same as Figure \ref{fig:NGC2273_HI_maps}, but for NGC 6323.}
	\label{fig:NGC6323_HI_maps}
\end{figure*}

Figures \ref{fig:NGC2273_HI_maps}, \ref{fig:ESO558-G009_HI_maps}, \ref{fig:NGC4258_HI_maps}, \ref{fig:NGC5765b_HI_maps}, and \ref{fig:NGC6323_HI_maps} show the masked HI data for NGC 2273, ESO 558-G009, NGC 4258, NGC 5765b, and NGC 6323, respectively.  The ``moment 0" maps show the integral over velocity of the fitted Gaussian function at every spatial pixel, the ``spectral line peak amplitude" maps show the best-fit Gaussian amplitude for the fitted function at every spatial pixel, and the ``spectral line central velocity" maps show the best-fit velocity centroid for the fitted Gaussian function at every spatial pixel.  We did not detect line emission from J0437+2456, CGCG 074-064, or NGC 6264.  The HI maps for NGC 1194, UGC 3789, and Mrk 1419 have been previously published in \cite{2013ApJ...778...47S}, and so we do not reproduce the images here.

Though in principle the line profile along any particular line of sight might not be well-fit by a Gaussian, this technique works well for visualization.  Our primary quantity of interest is the line-of-sight velocity at every spatial pixel, and in this regard the parametric fitting technique performs at least as well as a moment map approach (and often much better).  Furthermore, any deviation of the spectrum from a Gaussian shape will manifest as inflated uncertainties in the derived parameters, thereby increasing the likelihood of that pixel getting masked out.

Because of the clear systematic uncertainties present in assuming some parametric form for the line shape along any given line of sight, we note that the inputs to the tilted-ring fit described in \S\ref{HIFittingAndExploration} did not undergo any CLEANing or parametric fitting.  The analyses were instead performed directly on the ``dirty" data cubes.

\section{Measuring galaxy recession velocities using HI} \label{GalaxyRecessionVelocities}

Our goal is to measure the recession velocities of galaxies in a manner that is both independent of the SMBH velocity and which ideally can achieve an accuracy and precision comparable to the several km~s$^{-1}$ level of the maser measurements.  Neutral hydrogen seen in emission is an appealing candidate for a tracer that can fulfill these conditions, as it is almost always optically thin (thus tracing the full volume of the galaxy) and it doesn't suffer from reddening or extinction.  HI is also frequently present out to large ($\sim$tens of kpc) galactic radii, which for quiescent systems will allow it to trace the global dynamics of the galaxy well outside of the SMBH sphere of influence.  An accurate model of the global HI dynamics in a galaxy will necessarily incorporate its recession velocity, and so the construction of such a model allows us to make a precise measurement of that velocity.

To this end, we have analyzed spatially resolved HI observations for a number of galaxies (see Table \ref{tab:Observations}).  For several of these systems, the emission is either too weak (ESO 558-G009, Mrk 1419, NGC 6323) or too disordered (UGC 3789, NGC 5765b) to model it.  Instead, for those we have made nominal recession velocity measurements using a technique often applied to single-dish HI observations.  However, for three galaxies -- NGC 2273, NGC 4258, and NGC 1194 -- we have sufficiently high signal-to-noise data and the global HI dynamics are sufficiently orderly for a tilted-ring model to reasonably apply.  In this section, we describe the modeling and fitting techniques used to extract the galaxy recession velocities from the HI data.

\subsection{Generating the HI model} \label{GeneratingHIModel}

We fit the HI rotation curves using a tilted-ring model (e.g., \citealt{1989A&A...223...47B}, \citealt{2007A&A...468..731J}).  In this scheme, the gas in the galaxy is modeled as a series of concentric circular annuli, with each annulus (``ring") having an associated circular velocity and fixed HI surface density.  Since we do not view the rings perfectly face-on (i.e., they are ``tilted"), they appear as elliptical annuli on the sky.  The free parameters are the center position $(x_0,y_0)$, the systemic velocity $V_0$, the circular velocity $V_c(r)$ and gas surface density $\Sigma(r)$ at the orbital radius of the ring, and two angular parameters ($i,\phi)$ describing the inclination angle (defined to be the angle between the normal to the ring and the line of sight; takes on values between $0^{\circ}$ and $90^{\circ}$) and position angle (defined to be the angle measured counterclockwise from due north to the receding half of the major axis; takes on values between $0^{\circ}$ and $360^{\circ}$) respectively.  We allow $V_c$, $\Sigma$, $i$, and $\phi$ to vary between different rings, but $x_0$, $y_0$, and $V_0$ are global variables (i.e., they take on the same value for all rings).

For a galaxy disk divided into a discrete number $N$ of rings, we use the subscript $n \in \{ 1,\ldots,N \}$ to denote the parameters for a single ring (e.g., a ring at mean orbital radius $r_n$ would have associated inclination angle $i_n$ and position angle $\phi_n$).  Each ring is uniformly populated with a large number ($\sim$$10^5$) of point particles that carry non-gravitating HI mass (or equivalently HI flux); the actual number of particles scales with the ring's area so that the final model remains homoscedastic.  All particle masses are the same within a single ring, and the mass of each individual particle is set so that the total mass in the annulus $m_n$ is equal to the product of its area and surface density (i.e., $m_n = A_n \Sigma_n = \pi \left( r_n^2 - r_{n-1}^2 \right) \Sigma_n$, with $r_0 \equiv 0$).  We note that modeling the HI distribution in this way assumes either that the gas is optically thin, or that it resides in many small (but individually optically thick) clouds.

In the frame of the galaxy we denote the location of a particle within a ring using polar coordinates $(r,\theta)$, where $\theta$ is measured counterclockwise from the receding half of the major axis.  We relate the disk position $(r,\theta)$ to an on-sky location $(x,y)$ through

\begin{subequations}
\begin{eqnarray}
x & = & x_0 - r \cos(\theta) \sin(\phi_n) - r \sin(\theta) \cos(\phi_n) \cos(i_n) , \label{eqn:OnSkyX} \\
y & = & y_0 + r \cos(\theta) \cos(\phi_n) - r \sin(\theta) \sin(\phi_n) \cos(i_n) . \label{eqn:OnSkyY}
\end{eqnarray}
\end{subequations}

\noindent Every particle within a single ring is treated as having the same circular velocity $V_c(r_n)$, so that the observed (i.e., on-sky) velocity $V(r,\theta)$ can be expressed as

\begin{equation}
V(r,\theta) = V_0 + V_c(r_n) \sin(i_n) \cos(\theta) . \label{eqn:OnSkyVelocity}
\end{equation}

\noindent Within a single ring, each particle $(r,\theta)$ thus maps uniquely to a point $(x,y,V)$ in the data cube phase space.  Constructing a model cube from this cloud of particles is then simply a matter of binning them into 3D $(x,y,V)$ voxels, with bin boundaries set to match those of the observed cube.

\subsection{HI model fitting and parameter space exploration} \label{HIFittingAndExploration}

The input we used for the fitting procedure described in this section was a ``dirty" cube; that is, the UV data were transformed to the image plane, but no CLEANing was performed prior to performing the model fit.  Instead, the model cube produced at each MCMC iteration was convolved with the ``dirty beam" before being compared to the data cube.  The goal is to alleviate systematic uncertainties that might be introduced into the data cube during the somewhat subjective CLEANing process.

If we denote the value associated with each observed $(x,y,V)$ voxel in the data cube as $z_i$ (with associated uncertainty $\sigma_i$) and the corresponding modeled voxel values as $\zeta_i$, we can express the likelihood function as

\begin{equation}
\ln\left( \mathcal{L} \right) = - \frac{1}{2} \sum_i \left[ \frac{\left( z_i - \zeta_i \right)^2}{\sigma_i^2} + \ln\left( \sigma_i^2 \right) \right] . \label{eqn:LogLikelihood}
\end{equation}

\noindent Equation \ref{eqn:LogLikelihood} assumes that each data point $z_i$ deviates from the ``true" value $\zeta_i$ by only a Gaussian-distributed noise factor with variance $\sigma_i^2$.  For our purposes we assume that $\sigma_i$ is the same for all data points (so that we can denote it as $\sigma$), and we have estimated the value of $\sigma$ by computing the RMS in line-free channels of the data cube.  Because the correlated noise present in interferometric images can result in systematically underestimated uncertainties, we introduce a new fitted parameter $\alpha$ that scales $\sigma$ in the likelihood function,

\begin{equation}
\ln\left( \mathcal{L} \right) = - \frac{1}{2} \sum_i \left[ \frac{\left( z_i - \zeta_i \right)^2}{\alpha^2 \sigma^2} + \ln\left( \alpha^2 \sigma^2 \right) \right] . \label{eqn:LogLikelihoodScaledUncertainty}
\end{equation}

\noindent We then treat $\alpha$ as a nuisance parameter in the final fit.

Our final model has two fixed parameters: the number of rings $N$ and the outermost radius $r_{\text{max}}$.  The value of $r_{\text{max}}$ is chosen to lie near the outer edge of the visually obvious emission in the data cube, and $N$ is then set to be roughly equal to the number of resolution elements that fit within a radius $r_{\text{max}}$.  The model also has $4+4N$ adjustable parameters: the global parameters $V_0$, $x_0$, $y_0$, and $\alpha$, and the ring-specific parameters $V_c$, $\Sigma$, $i$, and $\phi$ (for each of the $N$ rings).  We use a flat (i.e., uniform) prior distribution for all modeled parameters, with ranges given in Table \ref{tab:FittingParameters}.  The posterior probability density is then computed as the product of the likelihood function and the prior distribution.

We performed a MCMC search of the parameter space, using the affine-invariant sampler \texttt{emcee} \citep{2013PASP..125..306F} to draw sample vectors from the posterior distribution.  This sampler employs a large number (we use $\sim$$10^3$ for our searches) of ``walkers" that simultaneously explore the parameter space, with each walker's knowledge of the whereabouts of the other walkers allowing it to efficiently navigate even heavily degenerate spaces.  For a detailed description of the algorithm, see \cite{goodman2010ensemble}.  We initialize each walker with parameter values that are randomly selected from the prior distributions.

\floattable
\begin{deluxetable}{lcccc}
\tablecolumns{4}
\tabletypesize{\scriptsize}
\tablecaption{Tilted-ring model parameter initializations\label{tab:FittingParameters}}
\tablehead{	&	\colhead{Prior range}	&	\colhead{Prior range}	&	\colhead{Prior range} \\
\colhead{Parameter}	&	\colhead{(NGC 2273)}	&	\colhead{(NGC 4258)}	&	\colhead{(NGC 1194)}}
\startdata
$V_0$							&	1700 -- 2000 km~s$^{-1}$	&	400 -- 500 km~s$^{-1}$				&	3950 -- 4250 km~s$^{-1}$			\\
$x_0$							&	$-30$ -- 30	arcsec				&	$-240$ -- 240	arcsec					&	$-30$ -- 30	arcsec						\\
$y_0$							&	$-30$ -- 30	arcsec				&	$-240$ -- 240	arcsec					&	$-30$ -- 30	arcsec						\\
$\alpha$					&	0 -- 10										&	0 -- 10												&	0 -- 10												\\
$V_{c,n}$					&	0 -- 500 km~s$^{-1}$			&	0 -- 500 km~s$^{-1}$					&	0 -- 500 km~s$^{-1}$					\\
$\Sigma_n$				&	0 -- 0.1 Jy~arcsec$^{-2}$	&	0 -- 0.1 Jy~arcsec$^{-2}$			&	0 -- 0.01 Jy~arcsec$^{-2}$		\\
$i_n$							&	0 -- $\frac{\pi}{2}$			&	0 -- $\frac{\pi}{2}$					&	0 -- $\frac{\pi}{2}$					\\
$\phi_n$					&	0 -- $\frac{\pi}{2}$			&	$\frac{3 \pi}{2}$ -- $2 \pi$	&	$\frac{3 \pi}{2}$ -- $2 \pi$	\\
\midrule
$N$								&	10												&	20														&	6															\\
$r_{\text{max}}$	&	160 arcsec								&	800 arcsec										&	140 arcsec										\\
\enddata
\tablecomments{\textit{Top}: Fitted parameters used in the tilted-ring model described in \S\ref{GeneratingHIModel}.  $V_0$ is the systemic velocity of the galaxy, $x_0$ and $y_0$ are the coordinates of the center (relative to the phase center of the observations; see Table \ref{tab:Observations}), $\alpha$ is an uncertainty-scaling parameter, $V_{c,n}$ is the circular velocity of the $n$th ring, $\Sigma_n$ is the surface brightness of the $n$th ring, $i_n$ is the inclination angle of the $n$th ring, and $\phi_n$ is the position angle of the $n$th ring.  All prior probabilities are uniform within the specified range and zero outside of it. \textit{Bottom}: Fixed parameters for the tilted-ring model.  $N$ is the number of rings and $r_{\text{max}}$ is the maximum radius used in the fit.}
\end{deluxetable}

\subsection{HI integrated intensity profiles} \label{HISingleDish}

For all of the galaxies in which we have detected HI, even those for which a full tilted-ring model is not appropriate (either because of low signal-to-noise or complicated dynamics), we can still use the spatially integrated line emission to make a measurement of the recession velocity.  Such velocity measurements using HI profiles are routinely made with single-dish radio telescopes, which typically don't have sufficient angular resolution for spatially resolved HI spectroscopy on most galaxies.

\begin{figure*}[t]
	\centering
		\includegraphics[width=1.00\textwidth]{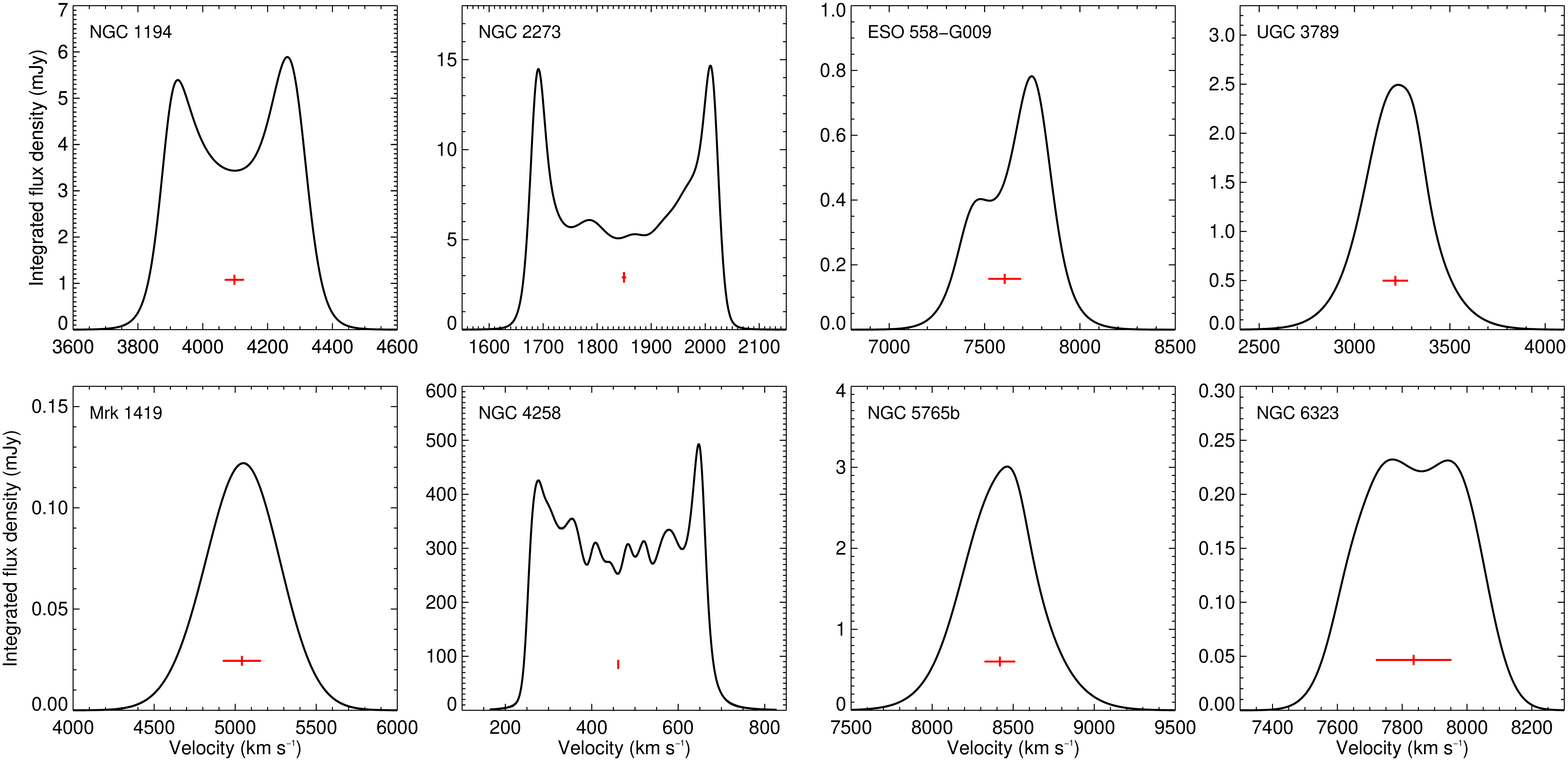}
	\caption{Spatially integrated HI profiles for all galaxies detected in HI, generated from the parameterized fits as described in \S\ref{HISingleDish}.  Note: these plots are noiseless because they represent fits to the data rather than the data themselves.  Measured $V_{20}$ velocities are marked as vertical red lines, and the associated uncertainties are shown as the horizontal red lines.  We note that the plot for NGC 5765b actually contains contributions to the HI profile from both NGC 5765b and its interacting companion, NGC 5765a, which is spatially blended with it in our VLA map.}
	\label{fig:HI_profiles}
\end{figure*}

Figure \ref{fig:HI_profiles} shows the integrated HI profiles for all detected galaxies in this work, generated using the masked parametric fitting results (see \S\ref{DataVisualization}).  These HI profiles are much cleaner than any that could actually be obtained from real single-dish observations.  The profiles are generated from summing many Gaussian fits to the data, which don't have noise.  We only considered emission within the masked region and are thus able to, e.g., isolate the emission of UGC 3789 from that of UGC 3797 (separated by $\sim$4.3 arcminutes).  We note that even with our VLA observations, the emission from NGC 5765b remains entangled with the emission from NGC 5765a (see Figure \ref{fig:NGC5765b_HI_maps}), so our integrated intensity profile actually contains contributions from both galaxies and therefore is not expected to provide a trustworthy recession velocity.

We use the method described in \cite{1990AAS...86..473F} to assign a recession velocity and uncertainty to each HI profile.  The recession velocity, denoted $V_{20}$, is defined to be to the midpoint between the two points on the profile that rise to 20\% of the peak amplitude.  The authors explored a variety of different line profile shapes, and they settled on a generic form for the uncertainty associated with the velocity as given by

\begin{equation}
\sigma_V = \frac{4}{S} \sqrt{\frac{1}{2} R \left( W_{20} - W_{50} \right)}, \label{eqn:VelocityUncertainty}
\end{equation}

\noindent where $S$ is the signal-to-noise ratio (defined to be the peak amplitude divided by the RMS), $R$ is the spectral channel spacing (in km~s$^{-1}$), and $W_{20}$ and $W_{50}$ are the widths of the profile at 20\% and 50\% of the peak intensity, respectively.  We note that this expression only accounts for the statistical uncertainty in the measurement of the velocity centroid.  Systematic deviations between the integrated profile's centroid and the true recession velocity of the galaxy, especially for asymmetric profiles, remain a possibility.

\section{Measuring SMBH velocites} \label{MeasuringSMBHVelocities}

All of the galaxies presented in this paper have previously published maser rotation curves, and several of them also have associated full disk models.  For the most part we simply quote these prior results, but for NGC 1194 and NGC 2273 we have elected to fit our own rotation curves to the VLBI data from \cite{2011ApJ...727...20K}.  We do so because the uncertainties on the rotation curve-derived velocities given by \cite{2011ApJ...727...20K} contain a systematic component that accounts for possible deviations of the SMBH location from some coordinate origin.  However, such a deviation can actually be incorporated directly into the model, thereby potentially decreasing the uncertainty on the measured velocity.  In this section we describe the maser disk model that we have used to measure the SMBH velocities in NGC 1194 and NGC 2273, using the VLBI data from \cite{2011ApJ...727...20K}.

\subsection{Method for fitting maser rotation curves} \label{FittingRotationCurves}

Our model assumes that the maser spots reside in a flat (i.e., unwarped), edge-on accretion disk that feels only the point-source gravitational potential from the central SMBH.  Each maser spot has a measured position $(x_i,y_i)$ and velocity $v_i$.

The uncertainties in the position measurements are associated with the VLBI beam used to make the maser map, resulting in correlated uncertainties between the desired coordinates of right ascension and declination.  For VLBI observations, we can describe the synthesized beam as having dimensions of $b \times a$ (Gaussian standard deviation, in mas$\times$mas) oriented at a position angle of $\theta$ (in degrees east of north).  Because this beam is Gaussian, we can construct a covariance matrix \textbf{S} for the uncertainties in the position of a point source using the method described in Appendix \ref{app:CovarianceMatrix}.  Doing so yields expressions for the (co)variances in terms of the beam parameters

\begin{subequations}
\begin{eqnarray}
\sigma_x^2 & = & a^2 \cos^2(\theta) + b^2 \sin^2(\theta) , \\
\sigma_{xy} & = & a^2 \sin(\theta) \cos(\theta) - b^2 \sin(\theta) \cos(\theta) , \\
\sigma_y^2 & = & a^2 \sin^2(\theta) + b^2 \cos^2(\theta) .
\end{eqnarray}
\end{subequations}

In practice, the uncertainty in the position of any particular maser spot is only proportional to the beam shape (see, e.g., \citealt{1997PASP..109..166C}), with the constant of proportionality $\gamma_i$ being inversely related to the signal-to-noise ratio $R_i$ as

\begin{equation}
\gamma_i = \frac{1}{2 R_i} . \label{eqn:GammaFactor}
\end{equation}

\noindent Equation \ref{eqn:GammaFactor} is only approximately true for low signal-to-noise observations, but we have determined empirically that it holds well for $R_i \gtrsim 7$.  Our expression for the covariance matrix \textbf{S}$_i$ for a single maser spot can then be written as

\begin{equation}
\text{\textbf{S}}_i = \gamma_i^2 \begin{pmatrix}
\sigma_x^2 & \sigma_{xy} \\
\sigma_{xy} & \sigma_y^2
\end{pmatrix} . \label{eqn:CovMatrix}
\end{equation}

\noindent The uncertainties in velocity measurements $v_i$ are assumed to be negligible.

We parameterize the maser disk's on-sky appearance as being a line with position angle $\phi$ and perpendicular distance from the origin $b_{\perp} \equiv b \cos(\phi)$, with thickness determined by a Gaussian variance $h^2$ in the direction perpendicular from the line.  This thickness parameter accounts for both any intrinsic scale height the masers in the disk might have, as well as real vertical scatter caused by unmodeled warping or inclination of the disk plane.  In addition to the orientation and thickness of the masing disk, the model also fits for the SMBH location $x_0$, velocity $v_0$, and Keplerian constant $k \equiv GM$ (which effectively gives the mass of the black hole, assuming the distance to the galaxy is known).  The value of $y_0$ (i.e., the SMBH location in the $y$-direction) is uniquely determined from $\phi$, $b_{\perp}$, and $x_0$, 

\begin{equation}
y_0 = x_0 \tan(\phi) + \frac{b_{\perp}}{\cos(\phi)} ,
\end{equation}

\noindent so it does not enter in as an additional free parameter.

Following \cite{2010arXiv1008.4686H}, the log likelihood for $(\phi, b_{\perp}, h)$ is given by

\begin{equation}
\ln(\mathcal{L}_1) = - \frac{1}{2} \sum_i \left[ \frac{\Delta_{\perp,i}^2}{\left( \Sigma_{\perp,i}^2 + h^2 \right)} + \ln\left( \Sigma_{\perp,i}^2 + h^2 \right) \right] . \label{eqn:ThickDiskL1}
\end{equation}

\noindent Here, $\Delta_{\perp,i}$ is the perpendicular displacement of a maser spot with position $(x_i,y_i)$ from the line,

\begin{equation}
\Delta_{\perp,i} = y_i \cos(\phi) - x_i \sin(\phi) - b_{\perp} ,
\end{equation}

\noindent and $\Sigma_{\perp,i}^2$ is the projection of \textbf{S}$_i$ onto the space perpendicular to the line,

\begin{equation}
\Sigma_{\perp,i}^2 = \gamma_i^2 \left[ \sigma_x^2 \sin^2(\phi) - 2 \sigma_{xy} \sin(\phi) \cos(\phi) + \sigma_y^2 \cos^2(\phi) \right].
\end{equation}

Similarly, we can write the projection of \textbf{S}$_i$ onto the line of the disk as

\begin{equation}
\Sigma_{\parallel,i}^2 = \gamma_i^2 \left[ \sigma_x^2 \cos^2(\phi) + 2 \sigma_{xy} \sin(\phi) \cos(\phi) + \sigma_y^2 \sin^2(\phi) \right] . \label{eqn:SigmaParallel}
\end{equation}

\noindent Equation \ref{eqn:SigmaParallel} effectively gives the uncertainty in the radial coordinate for a maser spot (i.e., it is the projection of $\text{\textbf{S}}$ onto the plane of the disk).  The Keplerian disk model predicts that a maser with observed velocity $v_i$ should lie at an orbital radius of

\begin{equation}
r_{\text{orb},i} = \frac{k}{\left( v_i - v_0 \right)^2} .
\end{equation}

\noindent This is to be compared with the measured orbital radius, which is given by

\begin{equation}
\Delta_{\parallel,i} = \sqrt{(x-x_0)^2 + (y-y_0)^2} ,
\end{equation}

\noindent where $x$ and $y$ are the projected coordinates of a maser spot observed at $(x_i,y_i)$ onto the disk, and are themselves given by

\begin{subequations}
\begin{eqnarray}
x & = & x_i \cos^2(\phi) + y_i \sin(\phi) \cos(\phi) - b_{\perp} \sin(\phi) , \\
y & = & x_i \sin(\phi) \cos(\phi) + y_i \sin^2(\phi) + b_{\perp} \cos(\phi) .
\end{eqnarray}
\end{subequations}

The likelihood for this parallel component of the model is then obtained by assuming that the observed orbital radii differ from the model orbital radii only by a Gaussian uncertainty with variance $\Sigma_{\parallel,i}^2$ in the radial direction (i.e., they deviate only by an uncertainty associated with the beam size).  Doing so yields

\begin{equation}
\ln(\mathcal{L}_2) = - \frac{1}{2} \sum_i \left[ \frac{\left( \Delta_{\parallel,i} - r_{\text{orb},i} \right)^2}{\Sigma_{\parallel,i}^2} + \ln\left( \Sigma_{\parallel,i}^2 \right) \right] . \label{eqn:ThickDiskL2}
\end{equation}

\noindent We can combine Equation \ref{eqn:ThickDiskL2} with Equation \ref{eqn:ThickDiskL1} to yield an overall likelihood of the model,

\begin{equation}
\ln(\mathcal{L}) = \ln(\mathcal{L}_1) + \ln(\mathcal{L}_2) .
\end{equation}

Our final maser disk model has six free parameters: $V_0$, $x_0$, $\phi$, $b_{\perp}$, $h$, and $k$.  We fit for these using the same MCMC procedure described in \S\ref{HIFittingAndExploration} for the HI model fitting, and we use flat priors for all parameters.

\section{Results} \label{Results}

\begin{figure*}[t]
	\centering
		\includegraphics[width=1.00\textwidth]{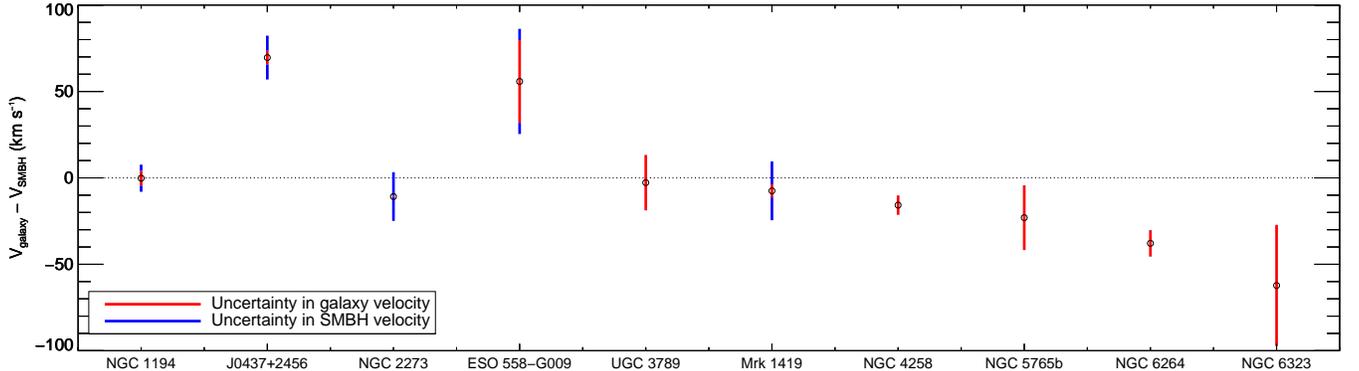}
	\caption{Differences between galaxy and SMBH systemic velocities.  The colors in the error bars show the quadrature contribution to the overall uncertainty from both measurements; red corresponds to the uncertainty contribution from the galaxy recession velocity measurement, and blue corresponds to that from the SMBH velocity measurement.  Two galaxies, J0437+2456 and NGC 6264, show statistically significant differences between the SMBH and host galaxy velocities.}
	\label{fig:velocity_comparison}
\end{figure*}

We have summarized the velocity measurements from this work, along with others from the literature, in Table \ref{tab:Velocity comparisons}.  The final measured differences between the galaxy and SMBH velocities are plotted in Figure \ref{fig:velocity_comparison}.  In this section we discuss details regarding the recession velocity and SMBH velocity measurements for J0437+2456 and NGC 6264, the two galaxies for which we see a statistically significant velocity offset.  The rest of the galaxies are discussed in Appendix \ref{app:IndividualDiscussion}.

\floattable
\begin{deluxetable}{lccc}
\tablecolumns{4}
\tabletypesize{\scriptsize}
\tablecaption{Fitting results\label{tab:FittingResults}}
\tablehead{\colhead{Parameter}	&	\colhead{NGC 2273}	&	\colhead{NGC 4258}	&	\colhead{NGC 1194}}
\startdata
$V_0$ (km~s$^{-1}$)						&	$1840.0_{-2.1}^{+2.4}$	&	$454.1_{-5.5}^{+5.6}$		&	$4088.6_{-5.6}^{+5.8}$	\\
$x_0$ (arcsec)								&	$0.0_{-5.0}^{+5.1}$			&	$-4.4 \pm 15.6$					&	$-2.2 \pm 5.9$					\\
$y_0$ (arcsec)								&	$-1.2_{-6.5}^{+6.9}$		&	$-47.8_{-21.4}^{+21.7}$	&	$-2.1_{-7.6}^{+7.7}$		\\
\midrule
$V_0$ (km~s$^{-1}$)						&	$1850.8_{-13.9}^{+13.5}$							&	\ldots	&	$4088.8 \pm 5.3$											\\
$x_0$ (mas)										&	$0.0553_{-0.0087}^{+0.0086}$					&	\ldots	&	$0.306_{-0.055}^{+0.056}$							\\
$\phi$ (degrees)							&	$332.97 \pm 0.42$											&	\ldots	&	$336.67_{-0.93}^{+0.95}$ 							\\
$b_{\perp}$ (mas)							&	$-0.0490 \pm 0.0032$									&	\ldots	&	$-0.237 \pm 0.053$ 										\\
$h$ (mas)											&	$0.0220_{-0.0029}^{+0.0031}$					&	\ldots	&	$0.167_{-0.033}^{+0.049}$							\\
$k$	(mas km$^2$ s$^{-2}$)			&	$2.680_{-0.029}^{+0.027} \times 10^5$	&	\ldots	&	$1.157_{-0.018}^{+0.017} \times 10^6$	\\
\enddata
\tablecomments{\textit{Top}: Results from fitting the HI tilted-ring model described in \S\ref{GeneratingHIModel} to NGC 2273, NGC 4258, and NGC 1194.  $V_0$ is the recession velocity of the galaxy, $x_0$ is the offset of the fitted center in right ascension from the phase center of the observations, and $y_0$ is the offset of the fitted center in declination from the phase center of the observations. \textit{Bottom}: Results from fitting the maser disk model described in \S\ref{FittingRotationCurves} to NGC 2273 and NGC 1194.  $V_0$ is the velocity of the SMBH, $x_0$ is its offset in RA from the phase center of the observations, $\phi$ is the position angle (measured east of north) of the edge-on accretion disk, $b_{\perp}$ is the perpendicular distance of the disk from the origin (i.e., from the phase center), $h$ is the Gaussian thickness of the disk, and $k$ is the Keplerian constant.  For all parameters in this table, the listed ``best fit" quantities are the 50th percentile values of the posterior distributions, with 1$\sigma$ uncertainties given as the 16th and 84th percentiles.}
\end{deluxetable}

\floattable
\begin{deluxetable}{lccccc}
\tablecolumns{6}
\tabletypesize{\tiny}
\tablecaption{Galaxy and SMBH recession velocities\label{tab:Velocity comparisons}}
\tablehead{	&	\colhead{$v_{\text{galaxy}}$}	&		&	\colhead{$v_{\text{SMBH}}$}	&		&	\colhead{Difference} \\
\colhead{Galaxy}	&	\colhead{(km~s$^{-1}$)}	&	\colhead{Method; citation}	&	\colhead{(km~s$^{-1}$)}	&	\colhead{Method; citation}	&	\colhead{(km~s$^{-1}$)}}
\startdata
\multirow{4}{*}{NGC 1194}	&	$\Big\{ 4088.6_{-5.6}^{+5.8} \Big\}$	&	HI tilted-ring fitting; this work										&	\multirow{4}{*}{$4088.8 \pm 5.3$}					&	\multirow{4}{*}{Maser rotation curve; this work}	&	\multirow{4}{*}{$-0.2 \pm 7.9$}	\\
													&	$4098 \pm 30$						&	HI integrated intensity profile; this work					&																						&		&		\\
													&	$4082.8 \pm 7.3$				&	Optical spectra; this work													&																						&		&		\\
													&	$4076 \pm 5$						&	HI single-dish profile; \cite{2005AA...430..373T}		&																						&		&		\\	\midrule
\multirow{3}{*}{J0437+2456}	&	$\Big\{ 4887.6 \pm 7.1 \Big\}$					&	Optical spectra; this work				&	\multirow{3}{*}{$4818.0 \pm 10.5$}				&	\multirow{3}{*}{Maser rotation curve; \cite{2017ApJ...834...52G}}	&	\multirow{3}{*}{$69.6 \pm 12.7$}	\\
													&	$4869.5 \pm 22.8$				&	Optical spectra; \cite{2012ApJS..199...26H}					&																						&		&		\\
													&	$4875.3 \pm 3.5$				&	Optical spectra; SDSS pipeline											&																						&		&		\\	\midrule
\multirow{4}{*}{NGC 2273}	&	$\Big\{ 1840.0_{-2.1}^{+2.4} \Big\}$	&	HI tilted-ring fitting; this work			&	\multirow{4}{*}{$1850.8_{-13.9}^{+13.5}$}	&	\multirow{4}{*}{Maser rotation curve; this work}	&	\multirow{4}{*}{$-10.8 \pm 14.1$}	\\
													&	$1850 \pm 4$						&	HI integrated intensity profile; this work					&																						&		&		\\
													&	$1839 \pm 4$						&	HI single-dish profile; \cite{1990AAS...82..391B}		&																						&		&		\\
													&	$1893 \pm 6$						&	Optical spectra; \cite{1995ApJS...99...67N}					&																						&		&		\\	\midrule
\multirow{2}{*}{ESO 558-G009}	&	$7606 \pm 86$				&	HI integrated intensity profile; this work					&	\multirow{2}{*}{$7618.2 \pm 14.0$}				&	\multirow{2}{*}{Maser rotation curve; \cite{2017ApJ...834...52G}}	&	\multirow{2}{*}{$55.8 \pm 30.4$}	\\
													&	$\Big\{ 7674 \pm 27 \Big\}$				&	Optical spectra; \cite{2012ApJS..199...26H}					&															&		&		\\	\midrule
\multirow{3}{*}{UGC 3789}	&	$3214 \pm 66$						&	HI integrated intensity profile; this work					&	\multirow{3}{*}{$3259.75 \pm 1.00$}				&	\multirow{3}{*}{Maser disk modeling; \cite{2013ApJ...767..154R}}	&	\multirow{3}{*}{$-2.8 \pm 16.0$}	\\
													&	$\Big\{ 3257 \pm 16 \Big\}$						&	Optical spectra; \cite{2012ApJS..199...26H}					&															&		&		\\
													&	$3325 \pm 24$						&	HI single-dish profile; \cite{1998AAS..130..333T}		&																						&		&		\\	\midrule
\multirow{2}{*}{Mrk 1419}	&	$5041 \pm 118$					&	HI integrated intensity profile; this work					&	\multirow{2}{*}{$4954.5 \pm 15$}					&	\multirow{2}{*}{Maser rotation curve; \cite{2011ApJ...727...20K}}	&	\multirow{3}{*}{$-7.5 \pm 17.0$}	\\
													&	$\Big\{ 4947 \pm 8 \Big\}$				&	HI single-dish profile; \cite{2005ApJS..160..149S}	&															&		&		\\	\midrule
\multirow{4}{*}{NGC 4258}	&	$\Big\{ 454.1_{-5.5}^{+5.6} \Big\}$		&	HI tilted-ring fitting; this work										&	\multirow{4}{*}{$466.87 \pm 0.49$}				&	\multirow{4}{*}{Maser disk modeling; \cite{2013ApJ...775...13H}}	&	\multirow{4}{*}{$-15.8 \pm 5.6$}	\\
													&	$461 \pm 0.3$						&	HI integrated intensity profile; this work					&																						&		&		\\
													&	$449 \pm 7$							&	HI single-dish profile; \cite{1981ApJS...47..139F}	&																						&		&		\\
													&	$443 \pm 3$							&	HI single-dish profile; \cite{1987MNRAS.224..953S}	&																						&		&		\\	\midrule
\multirow{3}{*}{NGC 5765b}&	$8418 \pm 95$						&	HI integrated intensity profile; this work					&	\multirow{3}{*}{$8322.22 \pm 1.13$}				&	\multirow{3}{*}{Maser disk modeling; \cite{2016ApJ...817..128G}}	&	\multirow{3}{*}{$-23.0 \pm 18.7$}	\\
													&	$\Big\{ 8299.2 \pm 18.7 \Big\}$			&	Optical spectra; this work							&																						&		&		\\
													&	$8329 \pm 30$						&	HI single-dish profile; \cite{2011AJ....142..170H}	&																						&		&		\\	\midrule
\multirow{3}{*}{NGC 6264}	&	$\Big\{ 10151.4 \pm 7.6 \Big\}$				&	Optical spectra; this work						&	\multirow{3}{*}{$10189.26 \pm 1.20$}			&	\multirow{3}{*}{Maser disk modeling; \cite{2013ApJ...767..155K}}	&	\multirow{3}{*}{$-37.9 \pm 7.7$}	\\
													&	$10177 \pm 28$					&	Optical spectra; \cite{1992BICDS..41...31H}					&																						&		&		\\
													&	$10161 \pm 76$					&	Optical spectra; \cite{2002AJ....123..100K}					&																						&		&		\\	\midrule
\multirow{2}{*}{NGC 6323}	&	$7835 \pm 117$					&	HI integrated intensity profile; this work					&	\multirow{2}{*}{$7834.28_{-2.2}^{+2.1}$}	&	\multirow{2}{*}{Maser disk modeling; \cite{2015ApJ...800...26K}}	&	\multirow{2}{*}{$-62.3 \pm 35.1$}	\\
													&	$\Big\{ 7772 \pm 35 \Big\}$				&	Optical spectra; \cite{1996AJ....112.1803M}					&															&		&		\\
\enddata
\tablecomments{Comparison of galaxy recession velocities ($v_{\text{galaxy}}$) and SMBH systemic velocities ($v_{\text{SMBH}}$) from the literature, along with the methods used to measure them; details for individual galaxies are given in \S\ref{Results}.  The listed galaxy recession velocities have been measured in this work using either HI tilted-ring fitting (\S\ref{HIFittingAndExploration}) or HI integrated intensity profile centroiding (\S\ref{HISingleDish}), and in other works using HI single-dish profile centroiding or optical spectral line fitting.  Velocities enclosed in curly brackets $\Big\{ \Big\}$ are those that we have used to compare with the SMBH velocities.  The listed SMBH velocities have been measured in this work using maser rotation curve modeling (\S\ref{FittingRotationCurves}), and in other works using either maser rotation curve modeling or full maser disk modeling.  The final velocity differences are given as $v_{\text{galaxy}} - v_{\text{SMBH}}$; only two galaxies, J0437+2456 and NGC 6264, show statistically significant differences between the galaxy and SMBH velocities.  All velocities are quoted in the barycentric reference frame and using the optical convention.}
\end{deluxetable}

\floattable
\begin{deluxetable}{lccccc}
\tablecolumns{6}
\tabletypesize{\scriptsize}
\tablecaption{Galaxy recession velocity measurements from SDSS spectra\label{tab:OpticalTransitions}}
\tablehead{	&		&	\multicolumn{4}{c}{Measured recession velocity (km~s$^{-1}$)} \\ \cline{3-6}
\colhead{Species}	&	\colhead{Rest wavelength ({\AA})}	&	\colhead{NGC 1194}	&	\colhead{J0437+2456}	&	\colhead{NGC 5765b}	&	\colhead{NGC 6264}}
\startdata
{[O II]}				&	3727.092				&	\ldots						&	$4814.3 \pm 77.3$	&	$8307.5 \pm 20.6$	&	$10070.3 \pm 28.2$	\\
{[O II]}				&	3729.875				&	\ldots						&	$4814.3 \pm 77.3$	&	$8307.5 \pm 20.6$	&	$10070.3 \pm 28.2$	\\
{[Ne III]}			&	3869.860				&	$4011.8 \pm 18.3$	&	$4836.1 \pm 14.5$	&	$8278.5 \pm 4.3$	&	$10122.4 \pm 4.2$		\\
H$\zeta$				&	3890.166				&	\ldots						&	\ldots						&	$8251.6 \pm 0.9$	&	$10140.6 \pm 1.8$		\\
{[Ne III]}			&	3968.590				&	\ldots						&	\ldots						&	$8278.5 \pm 4.3$	&	$10122.4 \pm 4.2$		\\
H$\varepsilon$	&	3971.198				&	\ldots						&	\ldots						&	$8251.6 \pm 0.9$	&	$10140.6 \pm 1.8$		\\
H$\delta$				&	4102.892				&	\ldots						&	\ldots						&	$8251.6 \pm 0.9$	&	$10140.6 \pm 1.8$		\\
H$\gamma$				&	4341.692				&	$4073.6 \pm 2.5$	&	\ldots						&	$8251.6 \pm 0.9$	&	$10140.6 \pm 1.8$		\\
{[O III]}				&	4364.435				&	\ldots						&	\ldots						&	$8295.7 \pm 0.8$	&	$10146.5 \pm 1.1$		\\
He II						&	4687.068				&	$4100.0 \pm 32.2$	&	\ldots						&	$8268.7 \pm 5.2$	&	$10143.8 \pm 6.9$		\\
{[Ar IV]}				&	4712.670				&	\ldots						&	\ldots						&	$8318.1 \pm 20.9$	&	$10158.6 \pm 17.3$	\\
{[Ar IV]}				&	4741.530				&	\ldots						&	\ldots						&	$8318.1 \pm 20.9$	&	$10158.6 \pm 17.3$	\\
H$\beta$				&	4862.691				&	$4073.6 \pm 2.5$	&	$4881.1 \pm 2.7$	&	$8251.6 \pm 0.9$	&	$10140.6 \pm 1.8$		\\
{[O III]}				&	4960.295				&	$4069.9 \pm 1.3$	&	$4887.1 \pm 1.5$	&	$8295.7 \pm 0.8$	&	$10146.5 \pm 1.1$		\\
{[O III]}				&	5008.240				&	$4069.9 \pm 1.3$	&	$4887.1 \pm 1.5$	&	$8295.7 \pm 0.8$	&	$10146.5 \pm 1.1$		\\
{[N I]}					&	5199.349				&	\ldots						&	\ldots						&	$8270.9 \pm 57.5$	&	$10152.6 \pm 43.9$	\\
{[N I]}					&	5201.705				&	\ldots						&	\ldots						&	$8270.9 \pm 57.5$	&	$10152.6 \pm 43.9$	\\
{[Fe VII]}			&	5722.300				&	\ldots						&	\ldots						&	$8311.1 \pm 11.4$	&	$10189.4 \pm 16.9$	\\
He I						&	5877.249				&	\ldots						&	\ldots						&	$8282.6 \pm 8.0$	&	$10144.6 \pm 14.6$	\\
{[Fe VII]}			&	6088.700				&	\ldots						&	\ldots						&	$8311.1 \pm 11.4$	&	$10189.4 \pm 16.9$	\\
{[O I]}					&	6302.046				&	$4067.5 \pm 16.9$	&	$4871.4 \pm 9.3$	&	$8264.7 \pm 5.2$	&	$10116.8 \pm 11.8$	\\
{[S III]}				&	6313.810				&	\ldots						&	\ldots						&	\ldots						&	$10127.6 \pm 28.7$	\\
{[O I]}					&	6365.535				&	\ldots						&	\ldots						&	$8264.7 \pm 5.2$	&	$10116.8 \pm 11.8$	\\
{[Fe X]}				&	6376.270				&	\ldots						&	\ldots						&	$8252.1 \pm 27.0$	&	\ldots							\\
{[N II]}				&	6549.860				&	$4081.2 \pm 4.2$	&	$4879.7 \pm 1.7$	&	$8259.2 \pm 1.0$	&	$10144.6 \pm 2.6$		\\
H$\alpha$				&	6564.632				&	$4073.6 \pm 2.5$	&	$4881.1 \pm 2.7$	&	$8251.6 \pm 0.9$	&	$10140.6 \pm 1.8$		\\
{[N II]}				&	6585.270				&	$4081.2 \pm 4.2$	&	$4879.7 \pm 1.7$	&	$8259.2 \pm 1.0$	&	$10144.6 \pm 2.6$		\\
He I						&	6679.995				&	\ldots						&	\ldots						&	$8282.6 \pm 8.0$	&	$10144.6 \pm 14.6$	\\
{[S II]}				&	6718.294				&	$4083.8 \pm 5.5$	&	$4869.0 \pm 4.0$	&	$8258.1 \pm 1.9$	&	$10146.1 \pm 2.9$		\\
{[S II]}				&	6732.674				&	$4083.8 \pm 5.5$	&	$4869.0 \pm 4.0$	&	$8258.1 \pm 1.9$	&	$10146.1 \pm 2.9$		\\
{[Ar III]}			&	7137.770				&	\ldots						&	\ldots						&	$8285.4 \pm 7.6$	&	$10145.3 \pm 7.7$		\\ \midrule
\textbf{Average (lines):}		&			&	$4071.7 \pm 8.1$	&	$4882.2 \pm 7.7$	&	$8271.1 \pm 23.2$	&	$10144.0 \pm 7.9$		\\
\textbf{Continuum:}					&			&	$4133.9 \pm 17.3$	&	$4921.4 \pm 19.1$	&	$8352.3 \pm 31.9$	&	$10234.8 \pm 26.5$	\\ \midrule
\textbf{Final:}					&			&	$4082.8 \pm 7.3$	&	$4887.6 \pm 7.1$	&	$8299.2 \pm 18.7$	&	$10151.4 \pm 7.6$ \\
\enddata
\tablecomments{A list of the optical emission lines used to measure redshifts from SDSS spectra.  We list the recession velocities as measured from each line species individually, along with the associated statistical uncertainty in the fit.  We also list the final velocity, which is a weighted average of the individual line velocities.  The uncertainty in the final velocity is a quadrature sum of the statistical uncertainty in the mean, the absolute calibration uncertainty of 2~km~s$^{-1}$ for SDSS spectra, and the magnitude of the scatter in velocities as measured using the different lines.  Rest wavelengths for each transition have been taken from the Atomic Spectra Database (ASD) provided by the National Institute of Standards and Technology (NIST).  In the second row from the bottom we list the velocities derived from the pPXF continuum fitting procedure, and the bottom row contains the final combined recession velocity measurements.}
\end{deluxetable}

\subsection{J0437+2456} \label{sec:J0437+2456}

No HI was detected in our VLA observations of J0437+2456.  Literature velocities for this galaxy include two optical spectra taken as part of the 2MASS Redshift Survey (2MRS; \citealt{2012ApJS..199...26H}) and three taken by SDSS.  Both 2MRS spectra were obtained in 2001 December with the FAST instrument onboard the 1.5-meter Tillinghast telescope at the Fred Lawrence Whipple Observatory, and the velocities were measured using the \texttt{XCSAO} package in IRAF \citep{1992ASPC...25..432K}, which employs a cross-correlation technique to compare observed spectra with templates.  The two 2MRS spectra have individual measured recession velocities of $4846.9 \pm 49.4$~km~s$^{-1}$ and $4874.5 \pm 23.1$~km~s$^{-1}$, for a combined velocity measurement of $4869.5 \pm 22.8$~km~s$^{-1}$.\footnote{The original velocity measurements for both of these spectra were $4835.8 \pm 23.6$~km~s$^{-1}$ and $4881.2 \pm 21.0$~km~s$^{-1}$, respectively.  However, Lucas Macri has generously re-run these spectra through the current incarnation of the 2MRS radial velocity code, and we report the updated velocity measurements in this paper.}  Of the three SDSS spectra, two were misclassified by the pipeline as being of either a star or a QSO.  Because the templates used by the SDSS pipeline to determine the recession velocity depend on the classification of the object (see \citealt{2012AJ....144..144B}), we have elected to make our own measurement of the recession velocity for J0437+2456 using the highest sensitivity SDSS spectrum available.\footnote{We note that for the one spectrum that the SDSS pipeline did correctly classify as an AGN, the listed recession velocity is $4875.3 \pm 3.5$~km~s$^{-1}$.  This small uncertainty is a result of the high signal-to-noise of the spectrum, and does not reflect the scatter observed when measuring velocities using individual emission lines in the spectrum (see Table \ref{tab:OpticalTransitions}); we thus retain our own velocity measurement, which incorporates this scatter into the quoted uncertainty.  Nevertheless, the SDSS pipeline measurement is consistent with both the 2MRS velocity and our own result.}  We used an optical spectrum (wavelength coverage $\sim$4000-10000~{\AA}) from the Baryon Oscillation Spectroscopic Survey (BOSS; \citealt{2013AJ....145...10D}) to measure the recession velocity of J0437+2456 to be $4887.6 \pm 7.1$~km~s$^{-1}$.  We have used the same fitting procedure and uncertainty calculations as for NGC 1194 (see Appendix \ref{sec:NGC1194}), and the results are listed in Table \ref{tab:OpticalTransitions}.

The SMBH velocity for J0437+2456 was measured by \cite{2017ApJ...834...52G}, who modeled the maser rotation curve using a method similar to that described in \S\ref{FittingRotationCurves}.  The authors used a two-step method to model the rotation curve, first measuring the disk's orientation on the sky and then rotating their coordinate system accordingly before performing the fit.  This approach avoids the need to include a position angle parameter in the fit, and though the authors assumed a thin disk (i.e., no parameter was included to account for potential disk thickness), they also included inclination angle as a fitted parameter.  Otherwise their method matches well with ours, and it yields a SMBH velocity of $4818.0 \pm 10.5$~km~s$^{-1}$.

We find a significant (5.5$\sigma$) difference between the galaxy and SMBH velocities for J0437+2456: the SMBH is blueshifted with respect to its host galaxy by $69.6 \pm 12.7$~km~s$^{-1}$.  If we consider the velocities derived from the stellar continuum and optical emission lines separately, we see a 4.7$\sigma$ and 4.9$\sigma$ difference, respectively, between the SMBH and galaxy velocities.  We can further subdivide the emission line measurements and consider only those expected to be the most reliable tracers of the galaxy systemic velocity.  \citealt{2005AJ....130..381B} claims that the low-ionization [O II], [N II], and [S II] lines provide the most reliable measurements, while the high-ionization [O III] lines are systematically blueshifted in a large fraction of galaxies.  Our velocity measurement of $4814.3 \pm 77.3$~km~s$^{-1}$ from the [O II] $\lambda$3727 doublet is consistent with no SMBH peculiar motion, but the uncertainty is large because the lines are so weak.  We measure more precise velocities of $4879.7 \pm 1.7$~km~s$^{-1}$, $4869.0 \pm 4.0$~km~s$^{-1}$, and $4887.1 \pm 1.5$~km~s$^{-1}$ for the [N II], [S II], and [O III] lines, respectively, all of which individually show significant deviations from the SMBH velocity (and we note that the [O III] lines do not display any systematic blueshift in J0437+2456).  A weighted average of the velocity measurements from only the low-ionization lines (including [O II]) returns $4878 \pm 1.6$~km~s$^{-1}$ (a 5.6$\sigma$ deviation from the SMBH velocity), and adding in the [O III] lines modifies the result to $4882.8 \pm 1.1$~km~s$^{-1}$ (a 6.1$\sigma$ deviation from the SMBH velocity).  The consistency between the stellar and gas velocities, and their common offset from the SMBH velocity, supports the interpretation of this system as a SMBH displaying peculiar motion.

It is possible that a misplacement of the SDSS fiber used to observe J0437+2456, or tracking drift in the center of that fiber during the course of the observation, could give rise to a systematic velocity offset that would mimic the signature of SMBH peculiar motion.  The listed central position of the fiber coincides with the PanSTARRS position to within several mas, and the worst-case image smearing induced by tracking errors in SDSS observations isn't expected to exceed $\sim$0.06~arcseconds \citep{2006AJ....131.2332G}.  We thus take 0.1~arcseconds to be a conservative upper limit to the magnitude of fiber displacement.  For a simulated rotating disk of material, a fiber with an aperture of 3~arcseonds that is misplaced from the galactic center by 0.1~arcseconds can pick up velocity offsets of up to $\sim$20~km~s$^{-1}$.  The exact value of the systematic velocity offset depends on a number of factors, including the direction of the fiber misplacement with respect to the symmetry axis of the rotating disk and the form of the rotation curve for the observed material.  Additional observations, ideally of spatially resolved gas and stellar kinematics (using, e.g., an integral field unit), will be necessary to provide a check against such systematics that can arise from fiber-fed spectroscopic measurements.

Assuming the measured velocity offset is real, it is natural to ask what the expected spatial separation of the SMBH from the galactic center might be.  Adopting a constant Milky Way-like central bulge mass density of $\rho \approx 190$~M$_{\odot}$~pc$^{-3}$ (see, e.g., \citealt{2013PASJ...65..118S}), we can estimate the maximum separation using

\begin{equation}
r_{\text{sep}} = \sqrt{\frac{3 (\Delta v)^2}{4 \pi G \rho}}, \label{eqn:Separation}
\end{equation}

\noindent where $\Delta v$ is the observed velocity difference between the SMBH and galaxy recession velocities.  For J0437+2456, we estimate a value of $r_{\text{sep}} \approx 37$~pc, which corresponds to approximately $0.1$~arcseconds on the sky.

The VLBI location of the maser system from \cite{2017ApJ...834...52G} is 04:37:03.6840 $+$24:56:06.837, with an uncertainty of 1.3~mas in right ascension and 2~mas in declination.  In Table \ref{tab:PositionalOffsets} we compare this SMBH position to the position of the galaxy as measured in three different astrometrically calibrated sky surveys: PanSTARRS, 2MASS, and AllWISE.  The galaxy positions from all three catalogs have a right ascension that is consistent with that of the SMBH, but both PanSTARRS and AllWISE show a statistically significant declination offset between the galaxy and SMBH.  The magnitude of this positional offset is $\sim$0.05~arcseconds in the optical (from PanSTARRS, detected at 3.3$\sigma$) and $\sim$0.25~arcseconds in the IR (from AllWISE, detected at 7.0$\sigma$), both of which match well with our rough prediction from the peculiar velocity.  In both cases, the SMBH appears to be offset to the south of the host galactic center.  We note that the PanSTARRS and AllWISE measurements are actually themselves statistically different from one another, perhaps indicating that the PanSTARRS position is being affected by extinction from dust within J0437+2456.

\floattable
\begin{deluxetable}{lccccc}
\tablecolumns{6}
\tabletypesize{\scriptsize}
\tablecaption{Positional offsets for J0437+2456 and NGC 6264\label{tab:PositionalOffsets}}
\tablehead{	&	&	\colhead{R.A.}	&	\colhead{Decl.}	&	\colhead{$\Delta_{\alpha}$}	&	\colhead{$\Delta_{\delta}$} \\
\colhead{Galaxy}	&	\colhead{Catalog}	&	\colhead{(J2000)}	&	\colhead{(J2000)}	&	\colhead{(mas)}	&	\colhead{(mas)}}
\startdata
\multirow{4}{*}{J0437+2456}	&	PanSTARRS						&	04:37:03.6830	&	$+$24:56:06.890	&	$-$$15 \pm 13$								&	$+$$53 \pm 16$							\\
														&	2MASS								&	04:37:03.6852	&	$+$24:56:06.918	&	$+$$18 \pm 70$								&	$+$$81 \pm 60$							\\
														&	AllWISE							&	04:37:03.6870	&	$+$24:56:07.090	&	$+$$45 \pm 39$								&	$+$$253 \pm 36$\hphantom{0}	\\ \cline{2-6}
														&	\textbf{Average}		&	04:37:03.6835	&	$+$24:56:06.923	&	\hphantom{0}$-$$8 \pm 12$			&	$+$$86 \pm 14$							\\ \midrule
\multirow{4}{*}{NGC 6264}		&	PanSTARRS						&	16:57:16.1280	&	$+$27:50:58.560	&	\hphantom{0}$+$$3 \pm 15$			&	$-$$17 \pm 9$\hphantom{0}		\\
														&	2MASS								&	16:57:16.1244	&	$+$27:50:58.657	&	$-$$51 \pm 90$								&	$+$$80 \pm 80$							\\
														&	AllWISE							&	16:57:16.1318	&	$+$27:50:58.539	&	$+$$60	\pm 36$								&	$-$$38 \pm 35$							\\ \cline{2-6}
														&	\textbf{Average}		&	16:57:16.1285	&	$+$27:50:58.560	&	$+$$10	\pm 14$								&	$-$$17 \pm 9$\hphantom{0}		\\
\enddata
\tablecomments{Positions for J0437+2456 and NGC 6264 taken from the PanSTARRS, 2MASS, and AllWISE catalogs.  The positional offsets in both right ascension ($\Delta_{\alpha}$) and declination ($\Delta_{\delta}$) are defined such that $\Delta \equiv P_{\text{galaxy}} - P_{\text{SMBH}}$, and the quoted 1$\sigma$ errors represent the uncertainty in the galaxy position, which dominates over the uncertainty in the SMBH position in all cases.}
\end{deluxetable}

\subsection{NGC 6264}

We did not detect HI emission in our VLA observations of NGC 6264.  \cite{1995AJ....109..874B} report a recession velocity of $10177 \pm 28$~km~s$^{-1}$, citing a private communication with Huchra, J.  An independent measurement is presented by \cite{2002AJ....123..100K}, who measured a recession velocity of $10161 \pm 76$~km~s$^{-1}$ for NGC 6264 using the same cross-correlation method as \cite{2012ApJS..199...26H}.  The authors added an uncertainty of 65~km~s$^{-1}$ in quadrature to that produced by the algorithm to account for possible systematic velocity offsets of the line-emitting region from the galaxy, increasing the uncertainty substantially over the pre-corrected value of 39~km~s$^{-1}$.

We have used an optical spectrum from SDSS \citep{2011AJ....142...72E} to measure the redshift of NGC 6264 via the same fitting procedure and uncertainty calculation employed for NGC 1194 (Appendix \ref{sec:NGC1194}).  The results from individual emission line and continuum fits are listed in Table \ref{tab:OpticalTransitions}, and our final recession velocity measurement is $10151.4 \pm 7.6$~km~s$^{-1}$.

The SMBH velocity for NGC 6264 ($10189.26 \pm 1.20$~km~s$^{-1}$) was determined by \cite{2013ApJ...767..155K}, who performed a full disk model of the maser system in this galaxy as part of the MCP.  The SMBH is redshifted with respect to its host galaxy by $37.9 \pm 7.7$~km~s$^{-1}$ (see Table \ref{tab:Velocity comparisons}).  However, this apparent offset is driven almost entirely by the optical emission line velocity measurement, which deviates from the SMBH velocity by 5.7$\sigma$.  The stellar continuum, by contrast, shows only a minor discrepancy with the SMBH velocity (they differ by 1.7$\sigma$), and in the opposite direction from that derived using the emission lines.  That is, the emission line velocity is blueshifted by 5.7$\sigma$ with respect to the SMBH velocity, while the continuum-derived velocity is redshifted by 1.7$\sigma$ with respect to the SMBH velocity.  It is thus plausible that the SMBH and stellar system share a common velocity while the optical emission lines trace gas that is blueshifted with respect to its host galaxy, possibly because they are preferentially tracing shocked gas (see, e.g., \citealt{2017arXiv170909177C}).

Nevertheless, if we proceed with the interpretation that the velocity offset between the SMBH and its host galaxy is caused by SMBH peculiar motion, then we can use Equation \ref{eqn:Separation} to estimate an expected spatial separation of $r_{\text{sep}} \approx 20$~pc, corresponding to 0.03~arcseconds at the distance to NGC 6264.  \cite{2011ApJ...727...20K} measured the VLBI maser position to be 16:57:16.1278 $+$27:50:58.5774, with an uncertainty of 0.3~mas in right ascension and 0.5~mas in declination.  We can see in Table \ref{tab:PositionalOffsets} that the position of NGC 6264 as measured by PanSTARRS, 2MASS, and AllWISE is consistent with both our estimate and with no separation.

\section{Conclusions} \label{Conclusions}

We have presented the idea of using H$_2$O megamasers in AGN accretion disks as dynamical tracers to measure SMBH peculiar motion, and we have applied this approach to a sample of galaxies for which VLBI data and maser rotation curves exist in the literature.  The galaxy recession velocities are measured using a combination of spatially resolved HI disk modeling, HI integrated intensity profile fitting, and optical spectral line and continuum fitting.

The ideal recession velocity measurement would perfectly reflect the motion of the galactic barycenter.  For galaxies with an undisturbed HI disk, spatially resolved modeling is a viable method for obtaining a precise measurement of the galaxy's recession velocity.  Optical spectra can be used to complement the HI measurements, and we have done so here where such spectra exist.

For two out of ten galaxies in our sample -- J0437+2456 and NGC 6264 -- we find a statistically significant ($>$3$\sigma$) difference between the SMBH velocity and its host galaxy's recession velocity.  In NGC 6264 the velocity of the stellar system matches that of the SMBH, and it seems likely that the apparent velocity offset between the optical emission lines and the SMBH arises from blueshifted ionized gas in the host galaxy, perhaps caused by AGN-driven shocks.  For J0437+2456, the velocity of the stellar system matches that of the optical emission lines, and both show a systematic redshift with respect to the SMBH velocity.  Furthermore, measurements of the galactic position from both PanSTARRS and AllWISE show statistically significant offsets (by roughly $\sim$0.1~arcseconds in declination) from the SMBH position, which match what we would expect given the magnitude of the measured velocity offset.  J0437+2456 is thus our most promising candidate for a true SMBH peculiar motion system.  We stress, however, that systematic effects arising from SDSS fiber misplacement can plausibly account for a large fraction of the observed velocity signal, and that additional observations will be necessary to corroborate the reality of the detected velocity offset.

If the spatial and kinematic offsets we see in J0437+2456 are genuinely tracing the SMBH motion, then the matching magnitude of the positional offset with the prediction from Equation \ref{eqn:Separation} is most easily explained by a solitary SMBH undergoing small-amplitude oscillations about the galactic center.  Such a scenario is expected in the aftermath of a binary SMBH merger, whereby the resulting post-merger SMBH experiences a kick that ejects it from the core of the galaxy.  Dynamical friction will quickly decay the SMBH orbit down to roughly the core radius, but beyond this point it ceases to operate as efficiently and the oscillations that occur on the core scale itself can last more than an order of magnitude longer than the initial decay timescale (see \citealt{2008ApJ...678..780G}).  J0437+2456 resides in a small group of galaxies \citep{2007ApJ...655..790C}, so it could plausibly have experienced the relatively recent galaxy merger (leading to a binary SMBH merger event) necessary for this interpretation to hold.

The observed velocity and positional offsets in J0437+2456 could also be explained if the SMBH is still in the process of inspiralling (i.e., post-galaxy merger but pre-SMBH merger; see, e.g., \citealt{2009ApJ...698..956C}).  The stellar mass interior to a $\sim$0.1~arcsecond orbital radius is expected to be roughly an order of magnitude larger than the mass of the SMBH itself, so the SMBH binary will not have hardened yet and the motion of the SMBHs will still be strongly influenced by the stellar potential.  Future observations should be able to place constraints on the presence of a possible companion SMBH.

We caution that both J0437+2456 and NGC 6264 have recession velocities measured only from optical emission lines and stellar continua in SDSS spectra, and that neither has corroborating HI measurements.  Future observations -- deeper HI spectra, spatially resolved optical spectroscopy, or both -- will be necessary to confirm whether these velocity offsets are real or whether the optically derived velocities are systematically shifted with respect to the galaxy's recession velocity.

For the remaining eight galaxies in our sample, five have SMBH peculiar velocity measurements that are currently limited by the precision in the host galaxy recession velocity, two are limited by the precision in the SMBH velocity, and one (NGC 1194) has comparable uncertainties in both measurements.  In cases where the SMBH velocity is the limiting factor, a complete maser disk model (as opposed to simply measuring the rotation curve) would substantially decrease the uncertainties.  For galaxies where the galactic recession velocity is the limiting factor, spatially resolved optical spectroscopy (such as provided by integral field units) will likely be a promising method to explore.

Peculiar velocity measurements for any single source are not by themselves enough to unambiguously identify the mechanism driving the motion.  However, making measurements of both positional and velocity offsets between a SMBH and its host galaxy, and/or making statistical measurements of velocity offsets for several sources, will allow us to narrow the range of possibilities.  Having a representative statistical sample of SMBH peculiar velocity measurements will help to constrain the efficiency of SMBH binary coalescence, a question that is becoming increasingly relevant as pulsar timing arrays push down the upper limit on a stochastic gravitational wave background.

\acknowledgments

We acknowledge Shane Davis, Aaron Evans, Jackie Huband, Remy Indebetouw, Cheng-Yu Kuo, Scott Suriano, Nick Troup, Mark Whittle, and Haifeng Yang for many helpful discussions, and Lucas Macri for re-running the 2MRS radial velocity code on their spectrum of J0437+2456.  We also thank the anonymous referee for comments that improved the quality of the paper.  Support for this work was provided by the NSF through the Grote Reber Fellowship Program administered by Associated Universities, Inc./National Radio Astronomy Observatory.  The National Radio Astronomy Observatory is a facility of the National Science Foundation operated under cooperative agreement by Associated Universities, Inc.  This research has made use of the NASA/IPAC Extragalactic Database (NED), which is operated by the Jet Propulsion Laboratory, California Institute of Technology, under contract with the National Aeronautics and Space Administration.

Funding for SDSS-III has been provided by the Alfred P. Sloan Foundation, the Participating Institutions, the National Science Foundation, and the U.S. Department of Energy Office of Science. The SDSS-III web site is http://www.sdss3.org/.

SDSS-III is managed by the Astrophysical Research Consortium for the Participating Institutions of the SDSS-III Collaboration including the University of Arizona, the Brazilian Participation Group, Brookhaven National Laboratory, Carnegie Mellon University, University of Florida, the French Participation Group, the German Participation Group, Harvard University, the Instituto de Astrofisica de Canarias, the Michigan State/Notre Dame/JINA Participation Group, Johns Hopkins University, Lawrence Berkeley National Laboratory, Max Planck Institute for Astrophysics, Max Planck Institute for Extraterrestrial Physics, New Mexico State University, New York University, Ohio State University, Pennsylvania State University, University of Portsmouth, Princeton University, the Spanish Participation Group, University of Tokyo, University of Utah, Vanderbilt University, University of Virginia, University of Washington, and Yale University.

The Pan-STARRS1 Surveys (PS1) and the PS1 public science archive have been made possible through contributions by the Institute for Astronomy, the University of Hawaii, the Pan-STARRS Project Office, the Max-Planck Society and its participating institutes, the Max Planck Institute for Astronomy, Heidelberg and the Max Planck Institute for Extraterrestrial Physics, Garching, The Johns Hopkins University, Durham University, the University of Edinburgh, the Queen's University Belfast, the Harvard-Smithsonian Center for Astrophysics, the Las Cumbres Observatory Global Telescope Network Incorporated, the National Central University of Taiwan, the Space Telescope Science Institute, the National Aeronautics and Space Administration under Grant No. NNX08AR22G issued through the Planetary Science Division of the NASA Science Mission Directorate, the National Science Foundation Grant No. AST-1238877, the University of Maryland, Eotvos Lorand University (ELTE), the Los Alamos National Laboratory, and the Gordon and Betty Moore Foundation.

This publication makes use of data products from the Two Micron All Sky Survey, which is a joint project of the University of Massachusetts and the Infrared Processing and Analysis Center/California Institute of Technology, funded by the National Aeronautics and Space Administration and the National Science Foundation.

This publication makes use of data products from the Wide-field Infrared Survey Explorer, which is a joint project of the University of California, Los Angeles, and the Jet Propulsion Laboratory/California Institute of Technology, and NEOWISE, which is a project of the Jet Propulsion Laboratory/California Institute of Technology. WISE and NEOWISE are funded by the National Aeronautics and Space Administration.

\facility{VLA}
\software{CASA}

\appendix

\section{Constructing the covariance matrix for a Gaussian beam} \label{app:CovarianceMatrix}

When determining the position of a point source as seen with an interferometer, the (relative) positional uncertainty along any direction is proportional to the beam size in that direction.  For a Gaussian synthesized beam, this uncertainty is fully characterized by a symmetric two-dimensional covariance matrix, \textbf{S}.  The diagonal elements of the covariance matrix correspond to the variances in the $x$ and $y$ directions (typically taken to be right ascension and declination), and the off-diagonal elements contain the covariance between $x$ and $y$ (i.e., information about the orientation of the Gaussian).  Writing \textbf{S} out in matrix form, we have

\begin{equation}
\text{\textbf{S}} = \begin{pmatrix}
\sigma_x^2 & \sigma_{xy} \\
\sigma_{xy} & \sigma_y^2
\end{pmatrix} . \label{eqn:CovarianceMatrix}
\end{equation}

If $\sigma_{xy}$ is zero then the beam is aligned with the coordinate axes, but in general the beam will have some dimensions $b \times a$ and a position angle $\theta$ that is misaligned in our coordinate system.  In this case, the beam can be described as a Gaussian with variances $a^2$ and $b^2$ that has been rotated by $\theta$ with respect to our coordinate system.  If we were to operate with \textbf{S} on a unit vector $\hat{\boldsymbol{v}}$ that points along one of the principle axes of the beam (let's say the $a$-axis), the only effect would be to scale the length of $\hat{\boldsymbol{v}}$ by the variance along that axis (i.e., $a^2$).  That is, $\hat{\boldsymbol{v}}$ is an eigenvector of \textbf{S} with eigenvalue $a^2$:

\begin{equation}
\text{\textbf{S}} \hat{\boldsymbol{v}} = a^2 \hat{\boldsymbol{v}} .
\end{equation}

\noindent An analogous expression holds for the vector pointed along the $b$-direction, and we can combine these two equations using the matrix expression

\begin{equation}
\text{\textbf{S}} \text{\textbf{V}} = \text{\textbf{V}} \text{\textbf{D}} . \label{eqn:CovarianceMatrix2}
\end{equation}

\noindent Here, \textbf{V} is the matrix whose columns are the eigenvectors of \textbf{S}, and \textbf{D} is the diagonal matrix whose elements are the corresponding eigenvalues of \textbf{S}.  \textbf{D} is thus the covariance matrix as viewed from a coordinate system that aligns with the principle axes of the beam, which we know is related to our coordinate system by nothing more than a rotation.  We can see that \textbf{V} must then be the rotation matrix that transforms from our coordinate system to the beam-aligned coordinate system, which we can re-denote as \textbf{R}.  Writing these two matrices out explicitly yields:

\begin{equation}
\text{\textbf{D}} = \begin{pmatrix}
a^2 & 0 \\
0 & b^2
\end{pmatrix} , \label{eqn:DiagonalMatrix}
\end{equation}

\begin{equation}
\text{\textbf{R}} = \begin{pmatrix}
\cos(\theta) & -\sin(\theta) \\
\sin(\theta) & \cos(\theta)
\end{pmatrix} . \label{eqn:RotationMatrix}
\end{equation}

\noindent Rearranging the terms in Equation \ref{eqn:CovarianceMatrix2} and substituting in \textbf{R} for \textbf{V} thus yields an expression for the covariance matrix in terms of the beam parameters:

\begin{equation}
\text{\textbf{S}} = \text{\textbf{R}} \text{\textbf{D}} \text{\textbf{R}}^{-1} . \label{eqn:CovarianceMatrix3}
\end{equation}

\section{Discussion of individual galaxies} \label{app:IndividualDiscussion}

Here we discuss the details of the velocity measurements for the 8 galaxies in our sample which showed no statistically significant velocity offset between the SMBH and host galaxy.

\subsection{NGC 1194} \label{sec:NGC1194}

\begin{figure*}[t]
	\centering
		\includegraphics[width=1.00\textwidth]{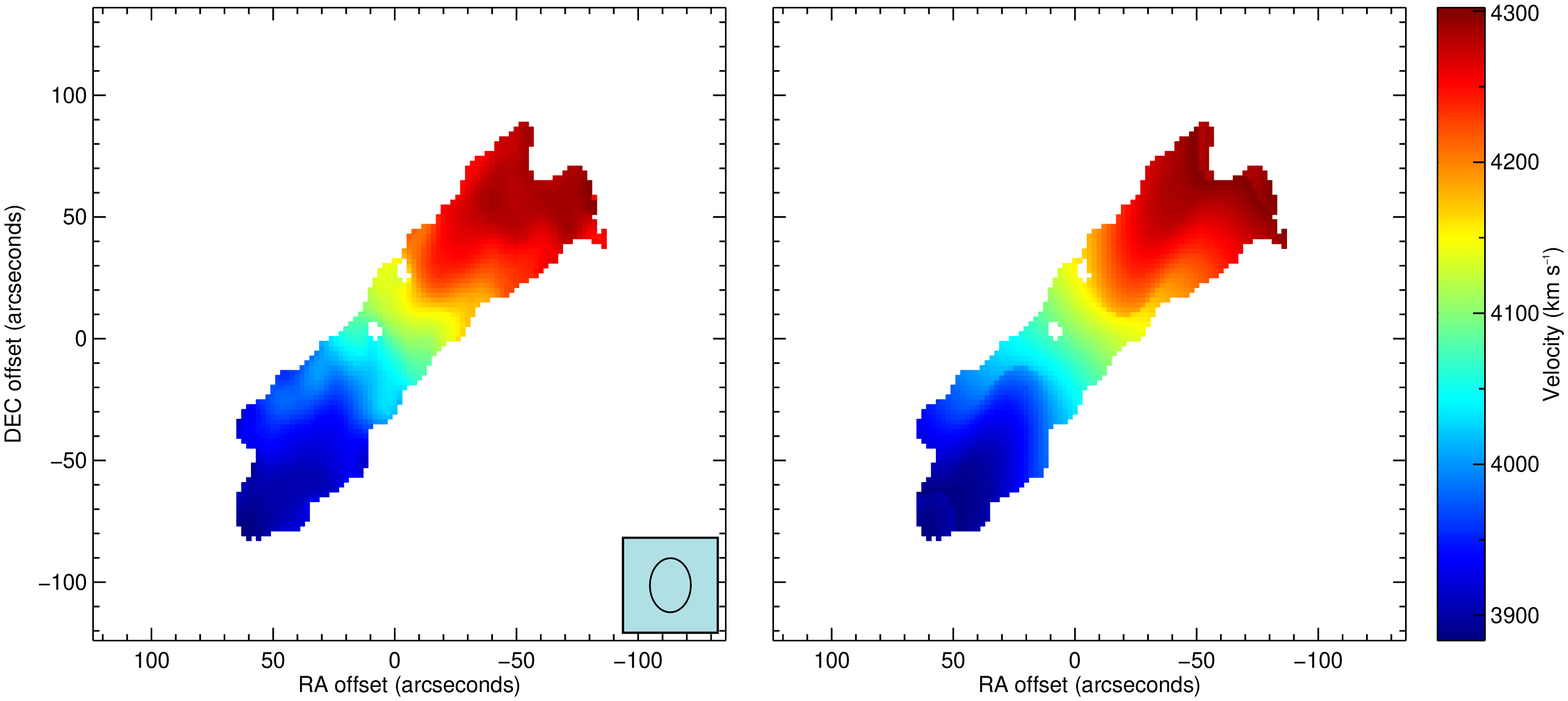}
	\caption{Observed (left) and modeled (right) velocity maps of the HI in NGC 1194, masked as described in \S\ref{DataVisualization}.  The coordinate axes mark the offset in right ascension and declination from the phase center of the observations (see Table \ref{tab:Observations}), and the half-power beam shape is shown in the bottom right-hand corner of the left plot.  The model velocity map has been constructed from the model cube and masked in the same manner as the data.}
	\label{fig:NGC1194_HI}
\end{figure*}

We fit a tilted-ring model to the HI disk in NGC 1194, using VLA data originally presented in \cite{2013ApJ...778...47S}; the results of the tilted-ring fitting are listed in the top portion of Table \ref{tab:FittingResults}, and the resulting velocity map is shown in Figure \ref{fig:NGC1194_HI}. The velocity we derive from the spatially integrated HI profile matches that obtained from the tilted-ring model, though the uncertainty is considerably larger in the former.  \cite{2005AA...430..373T} measured a single-dish HI velocity of $4076 \pm 5$~km~s$^{-1}$, consistent with our tilted-ring model fit of $4088.6_{-5.6}^{+5.8}$~km~s$^{-1}$.

We have also used an optical spectrum (wavelength coverage $\sim$4000-9000~{\AA}) from the Sloan Digital Sky Survey (SDSS; \citealt{2011AJ....142...72E}) to measure the recession velocity.  The spectral fitting was performed in two steps.  First, we fit the stellar continuum using the penalized pixel-fitting (pPXF) code developed by \cite{2017MNRAS.466..798C}, which fits for both the recession velocity and velocity dispersion using stellar population templates.  The stellar templates come from \cite{2004ApJ...613..898T}, who generated the templates using the simple stellar population models of \cite{2003MNRAS.344.1000B}.  All emission lines were masked out during continuum fitting.  We then subtracted this best-fit continuum model and fit Gaussians to all identified emission lines (listed in Table \ref{tab:OpticalTransitions}).  Each emitting species was constrained to have the same redshift (e.g., all [O III] lines have the same redshift, which may be different from that of the [O I] or [N II] lines), but otherwise all three Gaussian parameters (i.e., center, amplitude, and standard deviation) were independently initialized for each line within a wide, flat prior range.  Because several of the emission lines overlap -- most notably the [NII] and H$\alpha$ lines -- we performed simultaneous Gaussian fits to all of them.  Our resulting best-fit values for the recession velocity are listed in Table \ref{tab:OpticalTransitions}.

For the emission line velocity measurement, we consider three primary contributions to the uncertainty: a statistical uncertainty associated with the line-fitting, a systematic uncertainty associated with the absolute wavelength calibration of the spectrum (which is 2~km~s$^{-1}$ for SDSS spectra; \citealt{2009ApJS..182..543A}), and a systematic uncertainty arising from the choice of lines to fit.  The first two of these uncertainties are readily quantified, but the third is less clear.  In principle it is possible that some of the line emission (from, e.g., [OIII]) arises from gas with a systematically different dynamical behavior than that of the galactic barycenter.  To mitigate this source of uncertainty, we have incorporated the scatter (quantified as the standard deviation) between velocity measurements for the individual line species into the final uncertainty.  All sources of uncertainty were added in quadrature to arrive at the final value, which is listed in Table \ref{tab:OpticalTransitions}.

We estimate the systematic uncertainty in the continuum velocity measurement by making a series of separate pPXF fits to a sliding 500~{\AA} segment of the spectrum, and then calculating the $\chi^2$-weighted variance of all such fits.  This systematic uncertainty is then added in quadrature with the statistical uncertainty from the fit to the entire spectrum to arrive at the total uncertainty for the continuum-derived velocity measurement.  The final optical value for NGC 1194 ($4082.8 \pm 7.3$~km~s$^{-1}$) is taken to be the weighted mean of the fits from the continuum and from the emission lines.  This velocity measurement matches well with our HI tilted-ring model results.

\begin{figure*}[t]
	\centering
		\includegraphics[width=1.00\textwidth]{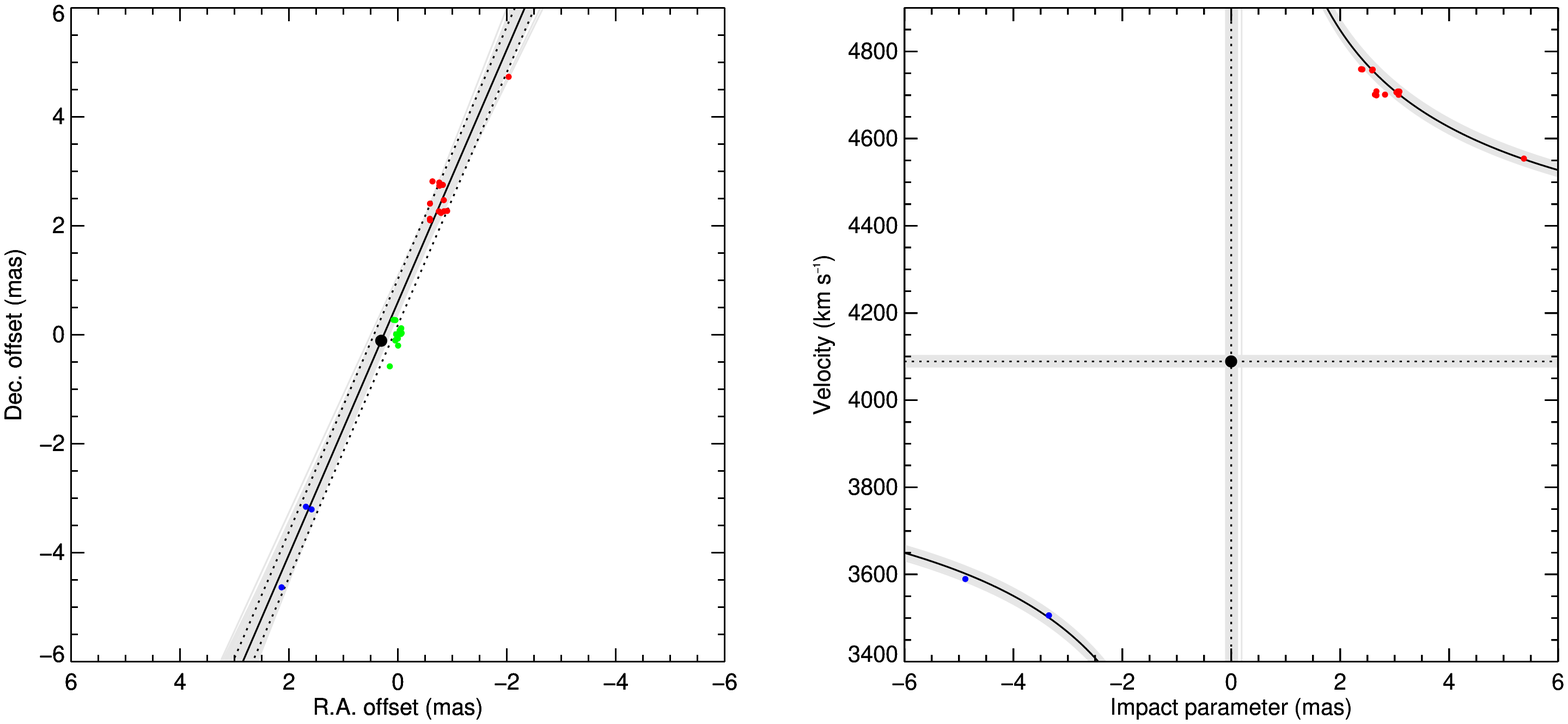}
	\caption{\textit{Left}: Map of the maser system in NGC 1194, with redshifted, systemic, and blueshifted maser spots plotted as red, green, and blue points, respectively.  The best-fit plane of the disk is plotted as a solid black line, and the dotted lines are offset by a perpendicular distance $h$ above and below the disk.  \textit{Right}: The best-fit rotation curve for NGC 1194 is plotted as a solid black line, and the best-fit velocity and dynamic center position are marked with horizontal and vertical dotted lines, respectively.  In both panels, the location of the dynamic center is plotted as a black point and the light grey lines show the fits from 100 different samplings of the posterior distribution.}
	\label{fig:NGC1194_map_rotcurve}
\end{figure*}

The bottom section of Table \ref{tab:FittingResults} contains the results from our maser disk model fit to NGC 1194, plotted in Figure \ref{fig:NGC1194_map_rotcurve}.  Our measured SMBH velocity is $4088.8 \pm 5.3$~km~s$^{-1}$, achieving an uncertainty smaller than the conservative 15~km~s$^{-1}$ value reported in \cite{2011ApJ...727...20K}.  We find no significant difference between the galaxy and SMBH velocities in NGC 1194.

\subsection{NGC 2273}

\begin{figure*}[t]
	\centering
		\includegraphics[width=1.00\textwidth]{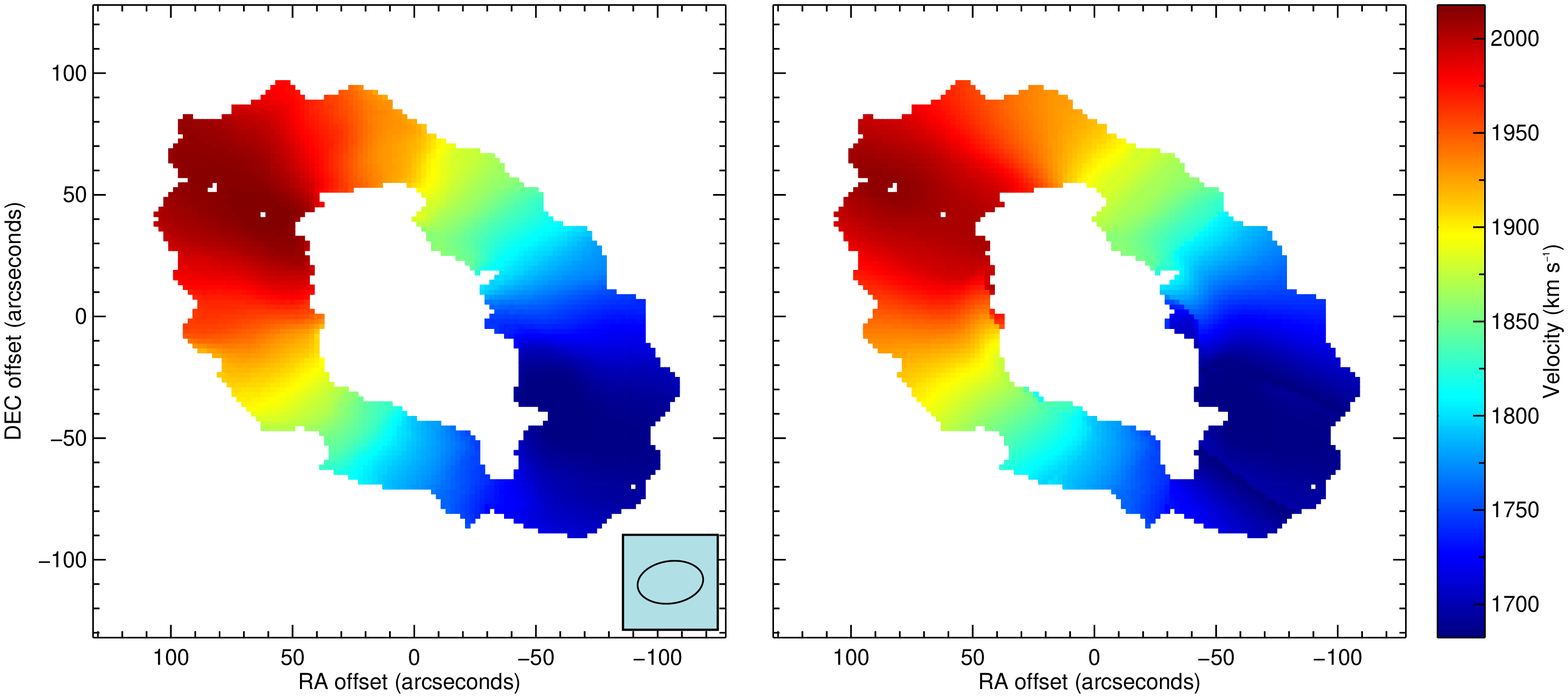}
	\caption{Same as Figure \ref{fig:NGC1194_HI}, but for NGC 2273}
	\label{fig:NGC2273_HI}
\end{figure*}

We used the VLA to observe HI in NGC 2273, and as with NGC 1194 (\S\ref{sec:NGC1194}) the HI disk is well-fit by a tilted-ring model.  The results of the tilted-ring fitting are listed in the top portion of Table \ref{tab:FittingResults}, and the resulting velocity map is shown in in Figure \ref{fig:NGC2273_HI}.  The data are well-fit by the model, and our final velocity measurement of $1840.0_{-2.1}^{+2.4}$~km~s$^{-1}$ has uncertainties consistent with the observed signal-to-noise ratio.  Though we also calculate the recession velocity from the spatially integrated HI profile, the uncertainty in that result is larger than what we obtain from the tilted-ring model.

Previous measurements of the recession velocity of NGC 2273 include both single-dish HI profile fitting and optical spectral line measurements.  \cite{1990AAS...82..391B} record the recession velocity as the midpoint of the HI line profile between the two points at 20\% of the peak value.  The quoted uncertainty depends on an empirically derived function of both the profile width and the spectral resolution (see the original paper for details), but their velocity of $1839 \pm 4$~km~s$^{-1}$ matches well with our result from the tilted-ring model.  \cite{1995ApJS...99...67N} took an optical spectrum of NGC 2273 and cross-correlated it with templates of stellar spectra -- a method developed by \cite{1979AJ.....84.1511T} -- to derive a recession velocity and associated uncertainty.  The authors recognized that their measured value of $1893 \pm 6$~km~s$^{-1}$ is considerably larger than the HI results, and they posit that this discrepancy might be caused by dust in the nuclear region of the galaxy.  We note that \cite{1995ApJS...99...67N} separately measured [O III] velocities and ``stellar velocities" (from the Ca II triplet and Mg \textit{b}); we only quote the stellar velocity in Table \ref{tab:Velocity comparisons}, but their measurement for [O III] alone is $1939 \pm 10$ km~s$^{-1}$ and thus even more discrepant from the HI values.  For our purposes, we retain the tilted-ring model result as the final recession velocity for NGC 2273.

\begin{figure*}[t]
	\centering
		\includegraphics[width=1.00\textwidth]{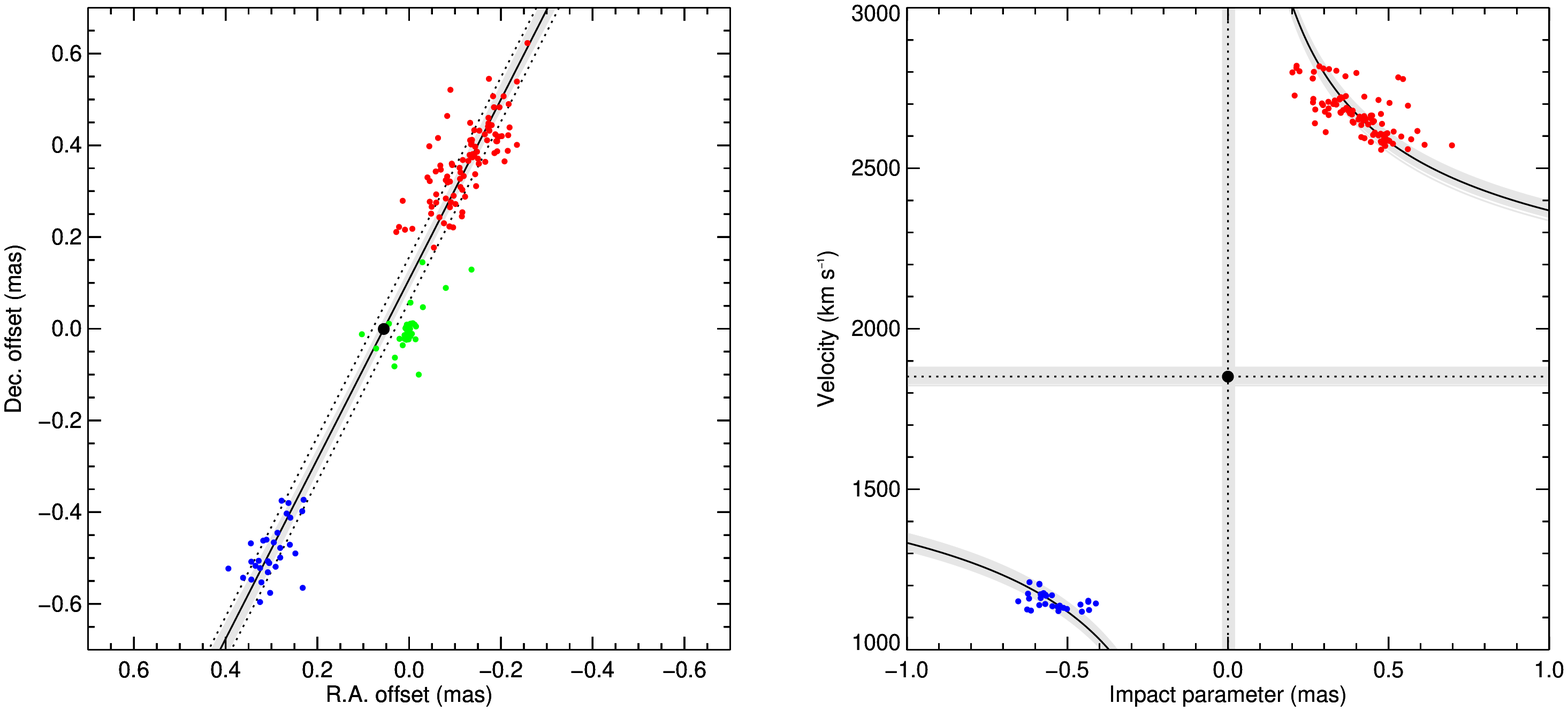}
	\caption{Same as Figure \ref{fig:NGC1194_map_rotcurve}, but for NGC 2273.}
	\label{fig:NGC2273_map_rotcurve}
\end{figure*}

We measured the SMBH velocity in NGC 2273 by modeling the rotation curve of the maser disk, and the results are listed in the bottom section of Table \ref{tab:FittingResults} and plotted in Figure \ref{fig:NGC2273_map_rotcurve}.  Our best-fit velocity is $1850.8_{-13.9}^{+13.5}$~km~s$^{-1}$.  As with NGC 1194, we determine a tighter constraint on the recession velocity than \cite{2011ApJ...727...20K}.  We find that there is no significant difference between the galaxy and SMBH velocities in NGC 2273.

\subsection{ESO 558-G009}

Though we detected HI in our VLA observations of ESO 558-G009 (see Figure \ref{fig:ESO558-G009_HI_maps}), the signal-to-noise ratio was insufficient to fit a tilted-ring model.  We did measure a recession velocity of $7606 \pm 86$~km~s$^{-1}$ using the spatially integrated HI profile, but the uncertainty in this measurement is large.  \cite{2012ApJS..199...26H} measured a recession velocity of $7674 \pm 27$~km~s$^{-1}$ using optical spectra, and we adopt their measurement.

As with J0437+2456 (\S\ref{sec:J0437+2456}), the SMBH velocity for ESO 558-G009 was measured by \cite{2017ApJ...834...52G} via maser rotation curve modeling.  We do not find a significant difference between the galaxy and SMBH velocities.

\subsection{UGC 3789}

We used the VLA observations of UGC 3789 from \cite{2013ApJ...778...47S} to measure a recession velocity from the spatially integrated HI profile, though as with ESO 558-G009 the resulting value of $3214 \pm 66$~km~s$^{-1}$ comes with a large uncertainty.  Single-dish HI measurements made by \cite{1998AAS..130..333T} using the NRT suffer from contamination by UGC 3797, located $\sim$4.3~arcminutes away.  \cite{2012ApJS..199...26H} have an optical measurement of $3325 \pm 24$~km~s$^{-1}$ for the recession velocity, which we adopt for this work.

The SMBH velocity for UGC 3789 has been precisely measured by \cite{2013ApJ...767..154R} as part of the Megamaser Cosmology Project (MCP).  The authors used multiple epochs of high-sensitivity VLBI and spectral monitoring to construct a geometric and kinematic model of the maser disk, constraining the SMBH velocity to a precision of $\sim$1~km~s$^{-1}$ in a manner that is fully independent of any galaxy-scale recession velocity measurements.  We find no significant difference between the galaxy and SMBH velocities in UGC 3789.

\subsection{Mrk 1419}

As with UGC 3789, we measure a recession velocity ($5041 \pm 118$~km~s$^{-1}$) from the spatially integrated HI profile of Mrk 1419 (using data taken by \citealt{2013ApJ...778...47S}), but the large uncertainty is insufficiently discriminating for our purposes.  \cite{2005ApJS..160..149S} measured a single-dish HI recession velocity of $4947 \pm 8$~km~s$^{-1}$ by fitting a first-order polynomial to the wings of the HI profile and using it to identify the 50\% flux level points on either side of the profile (i.e., the velocity at which the flux density reaches 50\% of the peak for the spectral horn on that side of the profile).  The recession velocity was then taken to be the average of these two velocities, and the authors determined the uncertainty in the velocity by simulating the observations using realistic noise and instrumental effects.  We adpot their measurement.

The SMBH velocity for Mrk 1419 was measured to be $4954.5 \pm 15$ by \cite{2011ApJ...727...20K} using maser rotation curve modeling.  There is no significant difference between the galaxy and SMBH velocities in Mrk 1419.

\subsection{NGC 4258}

\begin{figure*}[t]
	\centering
		\includegraphics[width=1.00\textwidth]{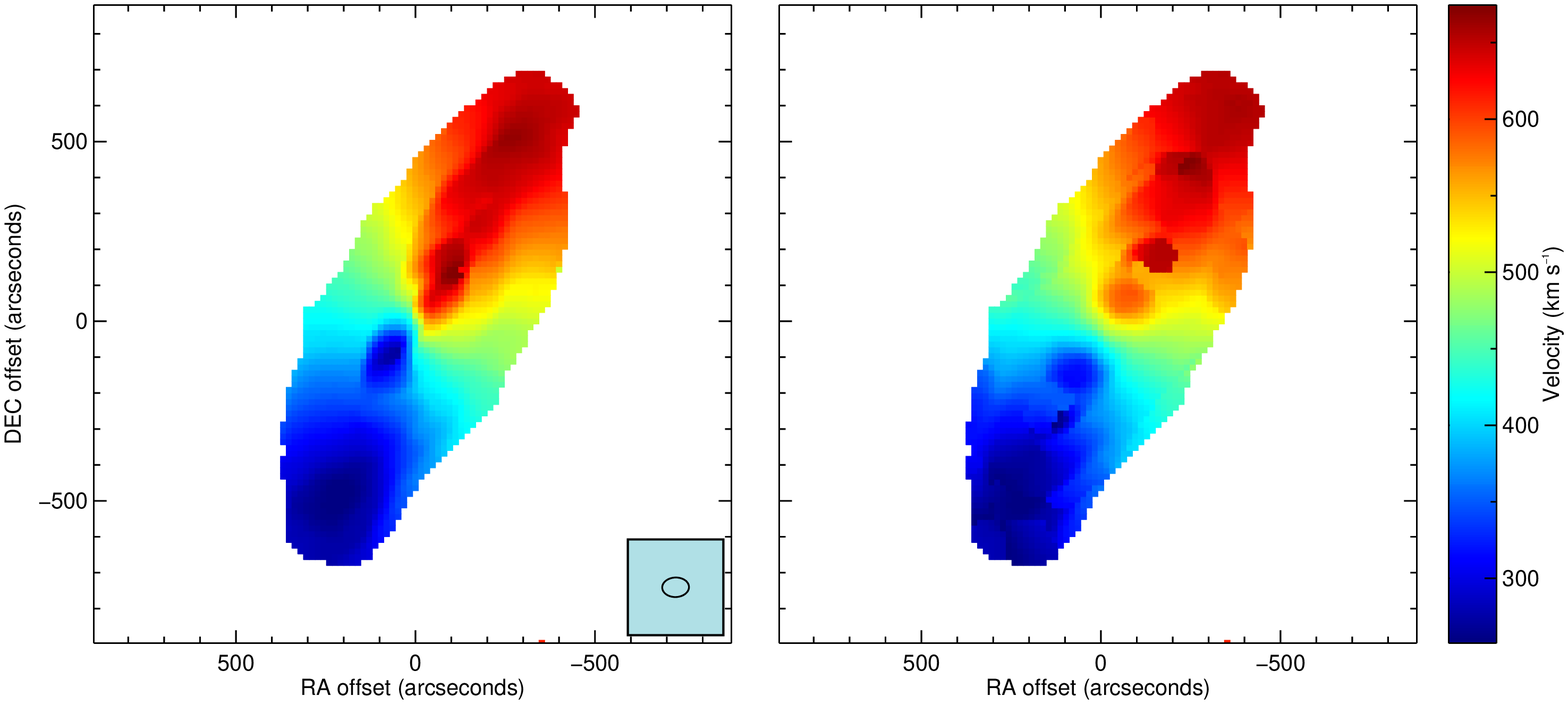}
	\caption{Same as Figure \ref{fig:NGC1194_HI}, but for NGC 4258.}
	\label{fig:NGC4258_HI}
\end{figure*}

We used archival VLA data to fit a tilted-ring model to the HI disk in NGC 4258; the results are listed in the top portion of Table \ref{tab:FittingResults}, and the resulting velocity map is shown in Figure \ref{fig:NGC4258_HI}.  Our best-fit recession velocity is $454.1_{-5.5}^{+5.6}$~km~s$^{-1}$.  The data are very high signal-to-noise, and there is evidence from the velocity map that the tilted-ring model fails to reproduce various dynamical structures in the disk; the uncertainty in the recession velocity for this galaxy is thus larger than what one might expect from consideration of signal-to-noise alone.  The velocity we derive from the spatially integrated HI profile matches that obtained from the tilted-ring model.  We consider the quoted uncertainty in the profile-derived recession velocity (obtained using Equation \ref{eqn:VelocityUncertainty}) to be an underestimate, because although it is driven down to small values by the extremely high signal-to-noise, it does not account for systematic deviations from an idealized HI profile.

NGC 4258 is a well-studied galaxy, and there are many examples of recession velocity measurements in the literature.  \cite{1987MNRAS.224..953S} used the 76-m telescope at Jodrell Bank to measure a HI recession velocity of $443 \pm 3$~km~s$^{-1}$ as the mean of the profile between the two points at 50\% of the peak flux value.  Though the uncertainty quoted in this measurement is small, it may be an underestimate for the same reasons we've described above regarding our spatially integrated HI velocity measurements.  An independent measurement by \cite{1981ApJS...47..139F}, using the 43-m radio telescope at Green Bank, obtained $449 \pm 7$~km~s$^{-1}$.  Both of these measurements are consistent with our result from the tilted-ring model fitting, which we retain as the recession velocity measurement for NGC 4258.

\cite{2013ApJ...775...13H} measured the velocity of the SMBH in NGC 4258 to be $466.87 \pm 0.49$ km~s$^{-1}$ via full-disk modeling of the maser system.  We have converted their value to the optical convention in the barycentric frame using

\begin{equation}
v_{\text{opt,bary}} = \left( \frac{v_{\text{rad,LSR}}}{1 - \frac{v_{\text{rad,LSR}}}{c}} \right) - 8.13 \text{ km s$^{-1}$} . \label{eqn:VelocityConversion}
\end{equation}

\noindent Here, $v_{\text{rad,LSR}}$ is the velocity measured using the radio convention in the LSR frame, and $v_{\text{opt,bary}}$ is the velocity measured using the optical convention in the barycentric frame.  Similar conversions have been made for other velocity measurements throughout this paper, when necessary.

We do not find a significant difference between the galaxy and SMBH velocities in NGC 4258.

\subsection{NGC 5765b}

We detected HI in our VLA observations of NGC 5765b, but the gas shows strong signs of kinematic disturbance (see Figure \ref{fig:NGC5765b_HI_maps}) from an interaction with the nearby (separation of $\sim$22 arcseconds) companion galaxy NGC 5765a.  The gas from both galaxies is spatially blended even in the VLA observations (Figure \ref{fig:NGC5765b_HI_maps}), which also show a large HI tail offset by $\sim$2~arcminutes from the optical center of either galaxy.  The integrated HI profile (Figure \ref{fig:HI_profiles}) for NGC 5765b is thus contaminated by emission from NGC 5765a, and so our velocity measurement derived from this profile is an unreliable tracer of the galaxy's motion.  This same issue holds true for the Arecibo HI spectrum of NGC 5765b measured by \cite{2011AJ....142..170H}.

We have thus used an optical spectrum from SDSS \citep{2011AJ....142...72E} to measure the recession velocity of $8299.2 \pm 18.7$~km~s$^{-1}$, using the same fitting procedure and uncertainty calculation as for NGC 1194 (\S\ref{sec:NGC1194}).  The results from individual emission line and continuum fits are listed in Table \ref{tab:OpticalTransitions}.

The SMBH velocity for NGC 5765b is $8322.22 \pm 1.13$~km~s$^{-1}$, from \cite{2016ApJ...817..128G}, who performed a full disk model of the maser system in this galaxy as part of the MCP.  We find no significant difference between the galaxy and SMBH velocities in NGC 5765b.

\subsection{NGC 6323}

We detected weak HI emission in our VLA observations of NGC 6323 (see Figure \ref{fig:NGC6323_HI_maps}), but it wasn't strong enough to fit a tilted-ring model.  The recession velocity of $7835 \pm 117$~km~s$^{-1}$ that we measured from the spatially integrated HI profile has a large uncertainty, also caused by the low signal-to-noise of the data.

An optical recession velocity measurement of $7772 \pm 35$~km~s$^{-1}$ was made by \cite{1996AJ....112.1803M}, who used the same template-matching methods as \cite{2012ApJS..199...26H}.  We adopt their value as the recession velocity for NGC 6323.

The SMBH velocity for NGC 6323 has been measured by \cite{2015ApJ...800...26K}, who performed a full disk model of the maser system in this galaxy as part of the MCP.  We do not find a significant difference between the galaxy and SMBH velocities in NGC 6323.

\bibliographystyle{apj}
\bibliography{mybib}

\end{document}